\pdfoutput=1 %arXiv admin added this line
\documentclass[12pt]{article}

\usepackage{latexsym,enumerate,natbib}
\usepackage{amsthm,amssymb,amsmath,amsfonts}
\usepackage{eucal,graphics}

\begin{document}

\bibliographystyle{plain}

\newcommand{\Hc}{{\mathcal H}}
\newcommand{\Om}{\Omega}
\newcommand{\e}{\ensuremath{\mathbb{E}}}
\newcommand{\1}[1]{\ensuremath{\mathbf{1}_{\{#1\}}}}
\newcommand{\p}{\ensuremath{\mathbb{P}}}
\newcommand{\hnorm}[2][H]{\ensuremath{\underline{\underline{#1}}^#2}}
\newcommand{\f}[1][F] {\ensuremath{\mathcal{#1}}}
\newcommand{\intt}[1][t] {\ensuremath{\int_0^{#1}}}
\newcommand{\inti} {\ensuremath{\int_0^\infty}}
\newcommand{\intinf} {\ensuremath{\int_{-\infty}^\infty}}
\newcommand{\intr} {\ensuremath{\int_\mathbb{R}}}
\newcommand{\vect}[1] {\ensuremath{\mathbf{#1}}}
\newcommand{\R}[1][R] {\ensuremath{\mathbb{#1}}}
\newcommand{\norm}[1] {\ensuremath{\Arrowvert #1 \Arrowvert}}
\newcommand{\stexp}[2][] {\ensuremath{\mathcal{E}_{#1}(#2)}}
\newcommand{\set}[1] {\ensuremath{\mathcal{#1}}}
\newcommand{\nn} {\ensuremath{\nonumber}}
\newcommand{\pd}[2] {\ensuremath{\frac{\partial{#1}}{\partial{#2}}}}
\newcommand{\ppd}[2] {\ensuremath{\frac{\partial^2 {#1}}{{\partial}{#2^2}}}}
\newcommand{\pdd}[3] {\ensuremath{\frac{\partial^2 {#1}}{{\partial}{#2}{\partial}{#3}}}}
\newcommand{\hf} {\ensuremath{\frac{1}{2}}}
\newcommand{\chf} {\ensuremath{\,_1F_1}}
\newcommand{\spa} {\ensuremath{\mbox{ }}}
\newcommand{\sgn}{\operatorname{sgn}}
\newcommand{\Lop}[1][L]{\ensuremath{\mathcal{#1}}}
\newcommand{\Lf}[1][L]{\ensuremath{\widehat{\mathcal{#1}}}}
\newcommand{\Ld}[1][L]{\ensuremath{\widetilde{\mathcal{#1}}}}
\newcommand{\dis}[1][t]{\ensuremath{e^{-\Lambda_{#1}}}}
\newcommand{\supp}{\operatorname{supp}}
\newcommand{\so}{\ensuremath{\Rightarrow\quad}}
\newcommand{\Lp}[1][g]{\ensuremath{(\mathcal{L} {#1})^+}}
\newcommand{\Lm}[1][g]{\ensuremath{(\mathcal{L} {#1})^-}}

\newcommand{\W}{\mathcal{W}}
\newcommand{\CI}{\mathcal{I}}
\newcommand{\SI}{{\mathcal{J}_{i}}}
\newcommand{\VIa}{\mathcal{V}^*}
\newcommand{\VI}{\mathcal{V}}
\newcommand{\LI}{\mathcal{A}}
\newcommand{\RI}{\mathcal{Z}}
\newcommand{\HI}{\mathcal{H}}
\newcommand{\ac}{\mathrm{ac}}
\newcommand{\s}{\mathrm{s}}
\newcommand{\Sr}{\mathcal{D}}
\newcommand{\Cr}{\mathcal{C}}
\newcommand{\op}{\mathrm{o}}
\newcommand{\cl}{\mathrm{c}}
\newcommand{\xl}{x_{l_i}}
\newcommand{\xr}{x_{r_i}}
\newcommand{\com}[1]{\textcolor{red}{#1}}

\newtheorem{defn}{Definition}[section]
\newtheorem{prop}{Proposition}[section]
\newtheorem{assm}{Assumption}[section]
\newtheorem{thrm}{Theorem}[section]
\newtheorem{cor}{Corollary}[section]
\newtheorem{lem}{Lemma}[section]
\newtheorem{exmp}{Example}[section]
\newtheorem{rem}{Remark}[section]
\newtheorem{chk}{Check!!}[section]

\title{The solution of discretionary stopping problems with applications to the optimal timing of investment decisions\footnote{Research supported by
EPSRC grant nos.\,GR/S22998/01,\, EP/C508882/1}\ \footnote{This is an Author's Original Manuscript of an article submitted for consideration in the IMA Journal of Mathematical Control and Information}}

\author{
{\sc Timothy C.\,Johnson\footnote{Maxwell Institute for
Mathematical Sciences and Department of Actuarial Mathematics
and Statistics, Heriot-Watt University, Edinburgh EH14 4AS,
UK, \texttt{t.c.johnson@hw.ac.uk}}}}

\maketitle

\begin{abstract}
We present a methodology for obtaining explicit solutions to infinite time horizon optimal stopping problems involving  general, one-dimensional, It\^o diffusions,  payoff functions that need not be smooth and state-dependent discounting.  This is done within a framework based on dynamic programming techniques employing variational inequalities and links to the probabilistic approaches employing $r$-excessive functions and martingale theory.  The aim of this paper is to facilitate the the solution of  a wide variety of problems, particularly in finance or economics.
\\\\
{\em Keywords\/}: Stochastic Control,  Optimal stopping, Dynamic programming, Finance
\\\\
{\em 2000 Mathematics Subject Classifications\/}:  60G40,
(93E20, 49K45, 91B70, 90C39)
\end{abstract}

\section{Introduction}\label{sec:intro}
A fundamental problem in finance, economics and  management science   is to determine the optimal time to invest in a project  in a random environment and to address these types of problems the theory of discretionary stopping has been widely employed following  Karlin~\cite{K_OMSA}.  

In order to solve some problems of this type we fix a filtered probability space, $(\Omega, \f, \f_t, \p)$, satisfying the usual conditions and carrying a standard one-dimensional $(\f_t)$-Brownian motion, $W$.  %We denote by $\CI$ a given open interval with left endpoint $-\infty \le \alpha$ and right endpoint $\beta  \le \infty$, and by ${\mathcal B}(\CI)$ the Borel $\sigma$-algebra on $\CI$. 
We assume that the stochastic system  we study is driven by the It\^{o} diffusion given by the stochastic differential equation 
\begin{align}
dX_t = b(X_t) \, dt + \sigma (X_t) \, dW_t , \quad X_0 = x \in \CI, \label{SDE}
\end{align} and the functions $ b  , \sigma : \CI \rightarrow \R$  satisfy Assumptions \ref{A1}--\ref{A2}. %and are such that (\ref{SDE}) has a weak solution ${\mathbb S}_x = (\Omega, \f, \f_t , \p_x, W , X)$ that is unique in the sense of probability law.% up to a possible explosion time, for all initial conditions $x \in \CI$.  We also assume that the solution of (\ref{SDE}) in non-explosive, i.e., the hitting time of the boundary $\{\alpha  , \beta  \}$ of the interval $\CI$ is infinite with probability 1 (see Karatzas and Shreve~\cite[Theorem~5.5.29]{ks1} for appropriate necessary and sufficient analytic conditions).

Our objective is to select the $(\f_t)$-stopping-time, $\tau$, that maximises 
\begin{align*}
  \e_x\Big[e^{-\Lambda_\tau}g(X_\tau ) \1{\tau<\infty} \Big], 
\end{align*}
where $g$ is subject to the conditions in Assumption \ref{A4} and $\Lambda$ is a state-dependent discounting factor defined by
\begin{equation}
\Lambda _t = \int _0^t r(X_s) \, ds , \label{Lam}
\end{equation}
for some  function $r$ satisfying the conditions of Assumption \ref{A3}.

The majority of financial and economic models in the literature assume that the underlying asset's value dynamics are modelled by a geometric Brownian motion, the associated payoff function is affine and the discounting rate is constant. The approach we employ relaxes all these assumptions.

Considering more general payoff functions allows for   utility based decision making, which, apart from the work of Henderson and Hobson~\cite{HenHob02}, and despite its fundamental importance, has not been widely discussed. 

However, the main benefit of accommodating general payoffs in the modelling framework  is the ability to incorporate compound payoffs into the payoff function, such as running payoffs or reversible decisions, as in Johnson and Zervos~\cite{JZ_ESSSP} or Guo and Tomecek~\cite{GT_CSCOS}. For example, consider the case where a project is initiated at a cost $G(X_t)$ and provides a running payoff given by $H(X_t)$.  In this case we have that
\begin{equation}
g(X_\rho ) =  -G(X_{\rho}) + \e_{X_\rho}\left[\int_{\rho}^{\infty}
  e^{-\Lambda_s}H(X_s)\,ds \right], \label{init}
\end{equation}
 for a stopping-time $\rho$.  Here, for example,  $X_t$ could represent the demand for electricity and $H$ is the `stack', a discontinuous function,  representing the price of supplying the demand, or $X_t$ could be the value of an asset that is taxed at banded rates, modelled by $G$ and $H$.  In both these, practically important, cases   $g$ will not be $C^2$ and the classical approach to solving stopping problems using variational inequalities cannot be used.

The framework we present accommodates systems  driven by general It\^o diffusions.  This is essential in economics given that not all asset price processes should be modelled by a geometric Brownian motion, which, on average, grows exponentially.  Beyond decisions driven by prices, extending the theory to a wider class of diffusions is important, for example in regime switching models such as in Dai, Zhang and Zhu~\cite{DZZ_TFTRSM} where the driving stochastic process represents a probability that the market is a bull or bear, and $\CI=]0,1[$.

Employing state dependent discounting enables a more realistic modelling framework for investment decisions in the presence of default risk and the events following 2007 have highlighted the importance of including this feature in decision making.

One approach to addressing discretionary stopping problems is through dynamic programming and involves  a set of variational inequalities.  This is discussed by, amongst many others,  El-Karoui~\cite{sf09elk},  Krylov~\cite{kryl80}, Bensoussan and Lions~\cite{BL_AVI},  Davis and Karatzas~\cite{DavKar94},  and Guo and Shepp~\cite{GuoSh01}. The technique  has become widespread in finance and economics since  the introduction  of so-called `real options' theory by McDonald and Siegel~\cite{McDS}, and has been described in Merton~\cite{M_CTF}, Dixit and Pindyck~\cite{DP94} and Trigorgis~\cite{T96}. However, taking this approach often involves making strong assumptions about the problem data in order to obtain explicit results.

A different approach has been to employ $r(\cdot)$-excessive functions, functions that satisfy
$$
f(x) \ge \e_x\left[e^{-\Lambda_\tau} f(X_\tau)\mathbf{1}_{\tau<\infty} \right],\quad\textrm{for }x\in\CI.
$$ 
This approach has been adopted by Dynkin~\cite{D_OC}, Shiryaev~\cite{shir78}, Salminen~\cite{Sal85}, Alvarez~\cite{A_RF} and Lempa~\cite{L_OSTS}, while  Dayanik and  Karatzas~\cite{daykar} and Dayanik~\cite{D_OSLDRD} use a certain characterisation of $r(\cdot)$-excessive functions, as the difference of two convex functions, to solve the stopping problem. These techniques, while powerful, are technical and are less accessible to a general audience in finance and economics unfamiliar with the details of probability theory.

Another approach uses martingale theory to locate the optimal boundaries between the {stopping}  and  {continuation} regions and is taken by, for example, Beibel and Lerche (\cite{bl97},~\cite{bl00}),   Lerche and  Urusov~\cite{LU_OSMT}, and Christensen and Irle~\cite{CI_HFTOS} and is described informally in Shreve~\cite[Sec 8.3.2]{S_SCF2}.  This approach, while relatively straightforward,  is based on assuming the diffusion starts in the continuation region and considers the first time it hits the stopping region.  However, as well as relying on the explicit problem data, this approach depends on the continuation region existing and an intuitive understanding of where it is located. 

The approach we adopt is based on the familiar dynamic programming approach while providing the power of the probabilistic techniques.  The  connection between the different approaches is provided by Johnson and Zervos~\cite{JZ1}, where the It\^o-Tanaka-Meyer formula is used to analyse the solution to the  variational inequality as the difference of two convex functions, rather than as a   function in $C^2$.  In using this result, the strong assumptions associated with the dynamic programming framework can be relaxed and explicit solutions that rely, only, on the problem data can be easily obtained. This approach has been taken in  R\"{u}schendorf and Urusov~\cite{RU_COSP}, Lamberton~\cite{L_OSIRF}, Johnson and Zervos~\cite{JZ_ESSSP} and has been developed fully in Lamberton and Zervos~\cite{LZ_OSODD}.  

The contribution of this paper is to demonstrate how the general theory, developed in  Lamberton and Zervos~\cite{LZ_OSODD}, can be applied to obtain explicit solutions to a variety of discretionary stopping problems.

This paper is organised as follows.  Section \ref{problem} is concerned with a formulation of the optimal stopping problem  and  a set of assumptions  for our problem to be well-posed while in Section \ref{implications} we discuss the practical implications of these assumptions.  
%We do this by describing first the solution technique in terms of  variational inequalities and the ``smooth-fit'' principle.  We then generalise this approach to a class of $C^1$ functions, using the results introduced in Section~\ref{problem}, and relate our approach to the Beibel and Lerche approach (\cite{bl97},~\cite{bl00}).  
In Section \ref{soln} we present the methodology for identifying the boundaries for six `elementary' problems and then, in Section \ref{ex}, we demonstrate how these `elementary' problems can be employed in solving more complex stopping problems.    An Appendix provides a proof of a key result in solving the problem when a continuation region lies between two stopping regions.

\section{Problem formulation and technical foundations} \label{problem}

%\begin{large} \end{large}In order to be able to accommodate the widest range of diffusions driving the system, we will adopt a weak formulation of the optimal stopping problem that we solve.
\begin{defn}\label{def:stop}
\rm{  Given an initial condition $x>0$, a \emph{stopping strategy} is any
  pair $(\mathbb{S}_x, \tau)$ such that
  $\mathbb{S}_x=(\Omega,\f,\f_t,\p_x,X,W)$ is a weak solution to  (\ref{SDE}) and $\tau$ is an $(\f_t)$-stopping-time.
   We denote by $\mathcal{S}_x$ the set of all such stopping
   strategies.
}\end{defn}

We consider the optimal stopping problem whose value function, $v$  is defined by
\begin{align}\label{def:v}
  v(x) = \sup_{(\mathbb{S}_x,
\tau)\in\mathcal{S}_x}J(\mathbb{S}_x, \tau),\quad \textrm{for }x \in \CI,
\end{align}
where
\begin{align*}
  J(\mathbb{S}_x, \tau)= \e_x\Big[e^{-\Lambda_\tau}g(X_\tau ) \1{\tau<\infty} \Big].
\end{align*}
Here, $g$ is the payoff function and $\Lambda$ a discounting function, discussed in Section \ref{sec:intro}. We shall now set out the Assumptions we require in order that this problem is well-posed.

To start this discussion, recall that in solving this problem through dynamic programming, we would associate the value function $v$ with a function $w$ that is the solution to the, so called, Hamilton-Jacobi-Bellman (HJB) equation,
\begin{align}\label{hjb}
  \max{\left\{
  \hf \sigma ^2 (x) w''(x) +  b (x) w'(x) - r(x) w(x),\ g(x)-w(x)\right\}} =0, \quad x\in \CI.
\end{align}
The solution to the stopping problem involves splitting the interval $\CI$  into  the {stopping } region,  $\Sr \subseteq\CI$, and the {continuation} region, $\Cr$,  with $\Cr=\CI/\Sr$. For all $x\in\,\Sr$, in order that (\ref{hjb}) is satisfied, we require that 
\begin{gather*}
g(x)-w(x) = \spa 0\quad\textrm{and}\quad\hf \sigma ^2 (x) w''(x) +  b (x) w'(x) - r(x) w(x) \le \spa 0, 
\end{gather*}
while  for all $x\in\,\Cr$,
\begin{gather*}
g(x)-w(x) \le \spa 0\quad\textrm{and}\quad\hf \sigma ^2 (x) w''(x) +  b (x) w'(x) - r(x) w(x) = \spa 0.
\end{gather*}

To develop an understanding of  the general solution of the second-order linear homogeneous ODE, which features  in (\ref{hjb}), we assume that the data of the one-dimensional It\^{o} diffusion, given by (\ref{SDE}) in the introduction, satisfies the following assumptions.
\begin{assm} \rm{
The functions $ b  , \sigma : \CI \rightarrow \R$ are ${\mathcal B} (\CI)$-measurable,
\begin{gather}
\sigma ^2 (x) > 0 , \quad \text{for all } x \in \CI, \nonumber \\
\intertext{and}
\int _{\underline{\alpha}}^{\overline{\beta }} \frac{1+ | b (s)|}
{\sigma ^2 (s)} \,ds < \infty \quad \text{and}  \quad \sup
_{s \in [\underline{\alpha}, \overline{\beta}]} \sigma^2 (s) <
\infty , \quad \text{for all } \alpha < \underline{\alpha} <
\overline{\beta} < \beta  . \nonumber
\end{gather}
\mbox{}\hfill$\Box$} \label{A1} \end{assm}
With reference to Karatzas and Shreve \cite[Section 5.5.C]{ks1}, the conditions appearing in this assumption are  sufficient for the SDE (\ref{SDE}) to have a weak solution ${\mathbb S}_x$ that is unique in the sense of probability law up to a possible explosion time, for all initial conditions $x \in \CI$.% In particular, given $c \in \CI$, the scale function $p_c$ and the speed measure $m$, given by
%\begin{gather}
%p_c (x) = \int _c^x \exp \left( -2 \int _c^s \frac{ b (u)}{\sigma^2(u)}
%\, du \right) ds, \quad \text{for } x \in \CI , \nonumber \\
%m (dx) = \frac{2}{\sigma ^2 (x) p' (x)} \, dx , \nonumber
%\end{gather}
%which characterise one-dimensional diffusions, are well-defined.
\begin{assm} {\rm
The solution of (\ref{SDE}) is non-explosive.
\mbox{}\hfill$\Box$} \label{A2} \end{assm}
This assumption means that the boundaries $\alpha$ and $\beta$ are inaccessible to the diffusion starting in $\CI$, though the boundaries can be entrance boundaries.

Relative to the discounting factor $\Lambda$, defined by (\ref{Lam}), we make the following assumptions.
\begin{assm} \rm{ The function $r: \CI \rightarrow \, ]0,\infty[$ is ${\mathcal B}
(\CI)$-measurable and there exists $r_0>0$ such that $r(x) \geq r_0$, for all $x \in
\CI$ and
\begin{gather}
\int _{\underline{\alpha}}^{\overline{\beta }} \frac{r(s)}
{\sigma ^2 (s)} \,ds < \infty, \quad \text{for all } \alpha < \underline{\alpha} <
\overline{\beta} < \beta  . \nonumber
\end{gather}
\mbox{}\hfill$\Box$}  \label{A3} \end{assm}

In the presence of Assumptions~\ref{A1}--\ref{A3} the general solution of the  ODE appearing  in (\ref{hjb}), 
\begin{equation}
\hf \sigma ^2 (x) f''(x) +  b (x) f'(x) - r(x) f(x) = 0 , \quad x \in
\CI, \label{H-ODE}
\end{equation}
 exists and is given by
\begin{equation}\label{GSODE}
f(x) = A \phi (x) + B \psi (x) ,
\end{equation}
for some constants $A, B \in \R$.

The functions $\phi$ and $\psi$ are $C^1$, their first derivatives are
absolutely continuous functions,
\begin{gather}
0 < \phi (x) \quad \text{and} \quad \phi ' (x) < 0 , \quad
\text{for all } x \in \CI ,  \label{phi-psi-prop1} \\
0 < \psi (x) \quad \text{and} \quad \psi ' (x) > 0 , \quad
\text{for all } x \in \CI , \label{phi-psi-prop2}\\
\intertext{and}
\lim _{x \downarrow \alpha } \phi (x) = \lim _{x \uparrow \beta }
\psi (x) = \infty . \label{phi-psi-lim}
\end{gather}
In this context, $\phi$ and $\psi$ are unique, modulo multiplicative
constants,  and the Wronskian, $\W$, is defined as
\begin{equation*}
\W (x) := \phi (x) \psi' (x) - \phi' (x) \psi (x) >0 , \quad
\text{for all } x,c \in \CI.
\end{equation*}

Also, given any points $x_1 < x_2$ in $\CI$ and weak solutions
${\mathbb S}_{x_1}$, ${\mathbb S}_{x_2}$ of the SDE
(\ref{SDE}), the functions $\phi$ and $\psi$ satisfy
\begin{equation}
\phi (x_2) = \phi (x_1) \e_{x_2} \left[ e^{-\Lambda _{\tau_{x_1}}}
\right] \quad \text{and} \quad
\psi (x_1) = \psi (x_2) \e_{x_1} \left[ e^{-\Lambda _{\tau_{x_2}}}
\right],  \label{phi-psi-prob}
\end{equation}
where $\tau_z$ denotes the first hitting time of $\{ z \}$, 
\begin{equation*}
\tau _z = \left\{ t \geq 0 \mid \ X_t = z \right\} .
\end{equation*}

All of these claims are standard and can be found in various forms in
several references, such as Feller~\cite{feller52}, Breiman~\cite{breiman}, It\^o and McKean \cite{itomck}, Karlin and Taylor~\cite{karlin},  Rogers and Williams \cite{r&w2} and Borodin and Salminen \cite{hndbkbm}.

In order to be able to address problems where the payoff $g$ is $C^1$, but not necessarily $C^2$, we will  consider signed measures of $\sigma$-finite total variation,   and we refer to them simply as ``measures''.  Given such a measure, $\mu$, on $(\CI, {\mathcal B} (\CI))$ we denote the total variation of $\mu$ by $|\mu| = \mu^+ + \mu^-$, where $\mu = \mu^+ - \mu^-$ is the Jordan decomposition of $\mu$.  Also, we say that a measure   is {\em non-atomic\/} if $\mu (\{c\}) = 0$, for all $c \in
{\CI}$.

A function $F: \CI\to \R$ is the difference of two convex functions if and only if its left-hand side derivative, $F'_-$, exists, is of finite variation, and its second distributional derivative is a measure, which we denote by  $F''(dx)$.  In this case we have the Lebesgue decomposition 
\begin{align*}%\label{eq:lebdecom}
F''(dx)=F_{\ac}''(x)dx + F_{\rm{s}}''(dx),
\end{align*}
where $F_{\ac}''(x)dx$ is absolutely continuous with respect to the Lebesgue measure and $F_{\rm{s}}''(dx)$ is mutually singular with the Lebesgue measure. On this basis, we define the measure $\Lop F$, related to the ODE (\ref{H-ODE}), by 
\begin{equation}
\Lop F (dx) = \hf \sigma ^2 (x) F'' (dx) + b(x) F_-'(x) \, dx - r(x)F(x) \, dx. \label{Lop}
\end{equation}

Now, we  define the finite variation, continuous processes $A^\mu$ by
\begin{equation*}
A_t^{\mu} = \int _{\alpha }^{\beta } \frac{L_t^y}{\sigma^2 (y)}
\, \mu (dy) , %\label{A}
\end{equation*}
where $L^y$ is the local-time process of $X$ at $y \in \CI$ and we have that the It\^o-Tanaka-Meyer formula gives  
\begin{equation*}%\label{ITM}
 e^{-\Lambda_t}F(X_t) = F(x) + \int_0^t e^{-\Lambda_u} dA_u^{\Lop F} + \int_0^t e^{-\Lambda_u} \sigma(X_u) F_-'(X_u) dW_u.
\end{equation*}
If $F$ is $C^1$ with a first derivative that is absolutely continuous with respect to the Lebesgue measure, $\Lop F(dx) = \Lop_{\ac} F(x) dx$, so that
\begin{equation*}%\label{Lac}
\int_0^t e^{-\Lambda_u} dA_u^{\Lop F} = \int_0^t e^{-\Lambda_u} \Lop_{\ac} F(X_u) du ,
\end{equation*}
we are able to  recover the familiar It\^o  formula from the It\^o-Tanaka-Meyer formula.

With this in mind and in order to place conditions on our payoff function, $g$, such that  our  problem is well-posed and does not involve value functions that are infinite, we introduce $(\phi,\psi)$-integrable measures.
\begin{defn}  {\rm
A measure $\mu$ on $(\CI , {\mathcal B} (\CI))$ is a $(\phi,\psi)$-integrable measure if
\begin{equation*}
\int _{]\alpha , \gamma [} \Psi (s) \, |\mu| (ds) + \int _{[\gamma ,
\beta [} \Phi (s) \, |\mu| (ds) < \infty , \quad \text{for all } \gamma
\in \CI ,
\end{equation*}
where the functions $\Phi$ and $\Psi$ are defined by
\begin{equation*}
\Phi (x) = \frac{\phi (x)}{\sigma ^2 (x) \W (x)} \quad \text{and}
\quad \Psi (x) = \frac{\psi (x)}{\sigma ^2 (x) \W (x)} .
%\label{Phi-Psi}
\end{equation*}
\mbox{}\hfill$\Box$} \label{d:int-meas} \end{defn}
Necessary and sufficient conditions for a measure $\mu$ on $(\CI , {\mathcal B} (\CI))$ to be $(\phi,\psi)$-integrable  are (\cite[Theorem 12]{LZ_OSODD})
\begin{gather}\label{mu-I}
\int _{\underline{\alpha}}^{\overline{\beta }}\frac{1}{\sigma^2(s)}|\mu|(ds) < \infty \quad\textrm{and}\quad \e_x\left[\int_0^\infty e^{-\Lambda_t} dA^{|\mu|}_t \right]<\infty
\end{gather}
for all $\alpha < \underline{\alpha} <\overline{\beta} < \beta$ and all $x\in\CI$.

On this basis, and with reference to  \cite[Theorem 12]{LZ_OSODD}, we make the following assumptions on $g$.
\begin{assm}{\rm
The function $g: \CI \rightarrow \R$ is  the difference of two convex functions, and the measure  $\Lop g$ is  $(\phi ,\psi)$-integrable.
In addition, $g$ has the following limiting behaviour
\begin{equation}
\lim _{x \downarrow \alpha} \frac{|g (x)|}{\phi (x)} =
\lim _{x \uparrow \beta} \frac{|g  (x)|}{\psi (x)} = 0 .
\label{limg}
\end{equation}
%\begin{equation}
%\lim _{x \downarrow \alpha} \frac{|R_\mu (x)|}{\phi (x)} =
%\lim _{x \uparrow \beta} \frac{|R_\mu  (x)|}{\psi (x)} = 0 .
%\label{limR}
%\end{equation}
\mbox{}\hfill$\Box$} \label{A4} \end{assm}

\section{Implications of the problem formulation} \label{implications}

The framework we adopt accommodates  the commonly encountered It\^o diffusions, including, Brownian motion, the Ornstein-Uhlenbeck process, geometric Brownian motion,  the geometric Ornstein-Uhlenbeck process and the so-called Feller, square-root mean-reverting, or Cox-Ingersoll-Ross, process defined by
\begin{gather*}%\label{SRMR}
  dX_t = \kappa(\theta-X_t)\,dt +\sigma\sqrt{X_t}\,dW_t, \qquad \CI=]0,\infty[,\quad X_0=x\in\ \CI,
\end{gather*}
where $\kappa$, $\theta$ and $\sigma$ are positive constants satisfying $\kappa\theta - \hf
\sigma^2 >0$.  This process  has an {entrance} boundary at $\alpha=0$, and $\lim_{x\downarrow\alpha} \psi(x)>0$.  When $r(x)=r$, the expressions for the general solutions (\ref{GSODE}) to the ODEs   associated with all of these diffusions  are all well known.  In situations where $\phi$ and $\psi$ are not known, it is possible to approximate them through simulation and employing (\ref{phi-psi-prob}).

Turning our attention to the payoff functions  that can be accommodated within our framework, we begin by observing that if $g$ is the difference of two convex functions and $|g|$ is bounded by some constant, it will be acceptable.  Similarly, if $g(y)=\infty$ for some $\alpha<y<\beta$ then (\ref{mu-I}) will fail.  In the case where $\lim_{x\downarrow\alpha}g(x) = \infty$ or $\lim_{x\to\beta}g(x) = \infty$, condition (\ref{limg}) of Assumption \ref{A4} is a strong test as to the acceptability of $g$.

For example, consider the case where $X$ is a geometric Brownian motion such that
$$
dX_t = bX_t + \sigma X_tdW_t,
$$
and suppose that $r(x)=r>0$, for constants $r, b$ and $\sigma$.  In this case it is well known that
\begin{gather*}
\phi(x) = x^m\quad\textrm{and}\quad\psi(x) = x^n
\end{gather*}
where $m<0<n$ are given by
$$
\left(\hf - \frac{b}{\sigma^2}\right) \pm \sqrt{\left(\hf - \frac{b}{\sigma^2}\right)^2 +\frac{2r}{\sigma^2}}.
$$
If $g(x)=x^j,\ j>0$, we need to establish what conditions on the problem data result in $g$ satisfying the conditions of Assumption \ref{A4}.
Noting that 
$$
\Lop g(dx) = \Lop_{\ac} g(x) dx = x^j\left(\hf\sigma^2(j(j-1)+ bj -r\right)dx,
$$
in order to establish the second condition in (\ref{mu-I}), we need to check the finiteness of 
\begin{align*}
   \e_x\left[\int_0^\infty  e^{-\Lambda_t} X_t^j\,dt\right]&=\\\
 % \e\left[\intt[\infty] x\exp\left\{\left(jb-j\frac{1}{2}\sigma^{2}+\frac{1}{2}j^2\sigma^{2}-r \right)t +
 %\left(-\frac{1}{2}j^2\sigma^{2}t+j\sigma W_{t} \right)\right\}  \,dt\right]\\
 %=&\spa
  \int_0^\infty &x\exp\left\{\left(jb-j\frac{1}{2}\sigma^{2}+\frac{1}{2}j^2\sigma^{2}-r \right)t\right\}
   \e\left[e^{\left(-\frac{1}{2}j^2\sigma^{2}t+j\sigma W_{t} \right)}\right]dt
 \end{align*}
 and $\Lop g$ is  $(\phi ,\psi)$-integrable if
 $$
 r >j b +  \hf j(j-1)\sigma^2.
 $$
This result can be obtained more directly by noting that
$$
 \frac{|g  (x)|}{\psi (x)} = \frac{x^j}{x^n}
$$
and so, to satisfy (\ref{limg}), we require that
$$
j<\left(\hf - \frac{b}{\sigma^2}\right) + \sqrt{\left(\hf - \frac{b}{\sigma^2}\right)^2 +\frac{2r}{\sigma^2}},
$$
which can be simplified to the previous inequality.

In fact, (\ref{limg}) will often be tautologous with the requirement that $\Lop g$ is $(\phi,\psi)$-integrable, however, the condition (\ref{limg}) is important in that it excludes the cases where $g(x)=\phi(x)$ or $g(x)=\psi(x)$.  

With this in mind,  under Assumptions  \ref{A1}--\ref{A4} and setting $$\mu(dx) = - \Lop g(dx),$$ the following results have been established in  Johnson and Zervos~\cite{JZ1},  Johnson and Zervos~\cite{JZ_ESSSP} or   in Lamberton and Zervos~\cite{LZ_OSODD}.

The payoff function $g$ can be expressed analytically as
\begin{align}
g (x) & = -\left(\phi (x) \int _{]\alpha , x[} \Psi(s) \, \Lop g (ds) +
\psi (x) \int _{[x , \beta [} \Phi(s) \, \Lop g (ds) \right) \nonumber \\
& \equiv -\left(\phi (x) \int _{]\alpha , x]} \Psi(s) \, \Lop g (ds) + \psi (x)
\int _{]x , \beta [} \Phi(s) \, \Lop g (ds)\right), \label{Rm-A}
\end{align}
 and probabilistically  as the $r(\cdot)$-potential of $A^{-\Lop g}$, specifically
\begin{equation*}%\label{Rm-P}
g(x) = \e_x \left[ \int _0^\infty e^{-\Lambda_t} \, dA_t^{-\Lop g}\right],
\end{equation*}
and it satisfies Dynkin's formula, i.e., given any
$(\f_t)$-stopping times $\rho_1 < \rho_2<\infty$,
\begin{align}
\e_x \left[ e^{-\Lambda _{\rho_2}} g (X_{\rho_2})
 \right] & =
\e_x \left[ e^{-\Lambda _{\rho_1}} g (X_{\rho_1})
 \right] +
\e_x \left[ \int _{\rho_1}^{\rho_2} e^{-\Lambda_t}
\, dA_t^{\Lop g} \right]. \label{DYNKIN}
\end{align}
In additiion we have a \emph{transversality} condition, namely,
given an increasing sequence of $(\f_t)$-stopping times $(\rho_n)$
such that $\lim _{n \rightarrow \infty} \rho_n = \infty$,
\begin{equation*}%\label{TVC}
\lim _{n \rightarrow \infty} \e_x \left[ e^{-\Lambda_{\rho_n}}
\left| g (X_{\rho_n}) \right| {\bf 1} _{\{ \rho_n
< \infty \}} \right] = 0. 
\end{equation*} 
This condition implies that our value function should be finite.

Furthermore, using (\ref{Rm-A}), we can calculate that
\begin{align}
g_+' (x)\phi (x)  - g(x)\phi' (x)  & =  -\W (x)
\int _{]x, \beta [} \Phi(s) \, \Lop g (ds), \label{e:R+phi} \\
g_-'(x)\phi (x)  - g(x)\phi' (x)  & =  -\W (x)
\int _{[x, \beta [} \Phi (s) \, \Lop g (ds), \label{e:R-phi} \\
g_+' (x)\psi (x)  - g (x)\psi' (x)  & = \W (x)
\int _{]\alpha, x]} \Psi (s) \, \Lop g (ds), \label{e:R+psi} \\
g_-' (x)\psi (x)  - g(x)\psi' (x)  & = \W (x)
\int _{]\alpha, x[} \Psi (s) \, \Lop g (ds). \label{e:R-psi}
\end{align}
Noting that
$$
\frac{d}{dx}\left(\frac{g(x)}{f(x)}\right)= \frac{g'(x)f(x) - g(x)f'(x)}{f^2(x)}
$$
we can see that (\ref{e:R+phi}--\ref{e:R-phi}) are related to the slope of the function $g/\phi$, while (\ref{e:R+psi}--\ref{e:R-psi}) relate to the slope of $g/\psi$.

Finally,  consider a function $H: \CI\to \R$ that is locally integrable with respect to the Lebesgue measure and define the measure $\mu^H$  on $(\CI, {\mathcal B} (\CI))$ by
$$
\mu^H(\Gamma) = \int_\Gamma H(x)\,dx,\qquad \Gamma\in {\mathcal B} (\CI).
$$
If $\mu^H$ is $(\phi ,\psi)$-integrable, then the function
$$
h(x) = \e_x\left[\,\int_0^\infty e^{-\Lambda_t} H(X_t)\, dt \right],\qquad \textrm{for } x \in \CI
$$
is $C^1$, has absolutely continuous first derivative and satisfies
$$
\Lop_{\ac} h(x) + H(x)=0.
$$
This result enables problems involving running payoffs, such as those in the initialisation example, (\ref{init}), to be tackled.

\section{The solution to six elementary stopping problems}\label{soln}

The methodology we employ is based on the results in Lamberton and Zervos \cite[Section 6]{LZ_OSODD}, where it is established that under Assumptions \ref{A1}--\ref{A3} and a weaker assumption on the payoff, that the value function, $v$,  associated with the  optimal stopping problem and defined by (\ref{def:v}),  is  of the form
\begin{align}\label{vsol}
 v(x) =&\spa \left\{\begin{array}{lcl}
    A\phi(x) + B \psi(x),& \textrm{if }  x\in\,\Cr\\
        g(x), & \textrm{if }   x\in\,\Sr,\\
  \end{array}\right.\quad\textrm{with }A,B \ge0,
\end{align} 
$v$ is $r(\cdot)$-excessive, and is a solution to the variational inequality
\begin{align}\label{e:hjb}
  \max{\left\{
  \Lop v(x),\ g(x)-v(x)\right\}}
  =0, \quad x\in \CI,
\end{align}
in the following sense.
\begin{defn} \rm{
A function $v:  \CI \mapsto \R$ is a solution of the variational inequality (\ref{e:hjb}) if $v(x)$ is the difference of two
convex functions,  the measure  $\Lop v$ is $ (\phi, \psi)$-integrable,
\begin{gather}
-\Lop v \text{ is a positive measure on } (\CI,
{\mathcal B}(\CI)) , \label{HJB1} \\
g(x)-v(x) \leq 0 ,
\quad \text{for all } x \in \CI , \label{HJB2} \\
\intertext{and }
\textrm{the measure }\Lop v\textrm{ does not charge the set } \Cr=\bigl\{ x \in \CI \mid v(x) >  g (x)\bigr\}. \label{HJB3}
\end{gather}
\mbox{}\hfill$\Box$}
\label{HJB-sense} \end{defn}
Since Lamberton and Zervos  consider only payoffs that are positive, in order to accommodate payoffs that are strictly negative, in particular for ${x\downarrow\alpha}$ and $x\uparrow\beta$, we need  the following growth condition on the value function
\begin{gather}
\bigl| v \bigr| \leq C \left( 1  + |g|\right) , \label{w-domin}
\end{gather}
for some constant $C>0$.  This condition, which is only relevant  for $x\in\Cr$, is found in the verification theorem in Johnson and Zervos \cite[Theorem 3]{JZ_ESSSP} and replaces (138) in the verification theorem in Lamberton and Zervos \cite[Theorem 13]{LZ_OSODD}.

This section of the paper presents a methodology for identifying the locations of the boundaries between the continuation region, $\Cr$, and the stopping region, $\Sr$, and as a consequence, the constants $A$ and $B$.  We shall consider six elementary cases in all; the cases are `elementary' in that by combining them together, more complex problems can be addressed.

We start  by observing that conditions (\ref{HJB1}) and (\ref{HJB3}) reveal immediately that if $\Lop g$ is \emph{positive} in some interval, then that interval \emph{cannot} be in the stopping region.  On this basis we have the first, most basic, two cases.

\begin{enumerate}[{CASE }I]
\item\label{GO} $\Lop g$ is {positive} for all $x\in\CI$. 
\item\label{STOP} $\Lop g$ is {negative} for all $x\in\CI$. 
\end{enumerate}

We now consider two more cases constructed by combining Cases \ref{GO} and Cases \ref{STOP}.  To appreciate the relevance of the first of these new cases, consider the situation when the payoff, $g$, is  such that $\lim_{x\downarrow\alpha}g(x)\le0$ and increases in such a way that there is a single boundary point, $x_\psi$, separating the continuation and stopping regions, so that $\Cr=]\alpha, x_\psi[$ and  $\Sr=[x_\psi,\beta[\,$.  In this situation, by (\ref{vsol}) and (\ref{w-domin}),  we must have $A=0$. 

In the standard approach, where $g$ is $C^2$, to specify the parameters $B$ and $x_\psi$, we would appeal to the so-called `smooth-pasting' condition of
optimal stopping that requires the value function to be $C^1$ at the free
boundary point $x_\psi$. This requirement yields the system of equations
\begin{align*}
B\psi(x_\psi) = g(x_\psi)\quad\textrm{ and }\quad B\psi'(x_\psi) = g'(x_\psi),
\end{align*}
which is equivalent to
\begin{align*}
B = \frac{g(x_\psi)}{\psi(x_\psi)}= \frac{g'(x_\psi)}{\psi'(x_\psi)}\quad\textrm{ and }\quad
q(x_\psi)=0,
\end{align*}
where $q$ is defined by
\begin{align*}
  q(x) := g(x)\psi'(x) - g'(x)\psi(x)\equiv \W (x)
\int _{\alpha}^{x} \Psi (s) \, \Lop g (ds),\quad x\in\CI,
\end{align*}
using (\ref{e:R+psi})--(\ref{e:R-psi}).  We note that $q(x_\psi)=0$ corresponds to a a stationary point of $g/\psi$. Since  $x_\psi\in\Sr$, and from (\ref{HJB1}) we require that $\Lop g\{x_\psi\}\le0$, we have the implication  that $x_\psi$ is at either a maximal turning point or a falling point of inflection.  

In the more general case that $g/\psi$ is not $C^1$ at the free boundary point $x_\psi$,  our objective is to find the parameter $B$ such that
\begin{align*}
B\psi'(x_\psi)\le &\spa g_-'(x_\psi)\\
B\psi(x_\psi)=&\spa g(x_\psi)\\
B\psi'(x_\psi)\ge &\spa g_+'(x_\psi).
\end{align*} Rearranging these using (\ref{e:R+psi})--(\ref{e:R-psi}) we have that
\begin{align*}
\int _{]\alpha, x_\psi]} \Psi (s) \, \Lop g (ds) \le 0 \le \int _{]\alpha, x_\psi[} \Psi (s) \, \Lop g (ds),
\end{align*}

At this point it is worth noting that the measure $\Lop g$ becomes central to obtaining the solution to the stopping problem, just as in Cases \ref{GO}--\ref{STOP}.  On this basis we present the next two cases. 
\begin{enumerate}[{CASE }I]
\setcounter{enumi}{2}
\item \label{CALL} $\Lop g$ is {positive} in the interval $]\alpha,x_r[$ and {negative} for all $x\in[x_r,\beta[$ such that there is a {maximal turning point of} $g/\psi$ for some $x_\psi$, with
$$
\frac{g(x_\psi)}{\psi(x_\psi)} \ge \frac{g(x)}{\psi(x)}, \qquad \textrm{for all } x\in\CI,
$$ 
but there is no maximal turning  point of $g/\phi$  in $\CI$.  This case is associated with call option type payoffs.
\item \label{PUT} $\Lop g$ is negative for all $x\in]\alpha,x_l]$ and positive in the interval $]x_l,\beta[$ such there is a maximal turning  point of $g/\phi$ for some $x_\phi$, with
$$
\frac{g(x_\phi)}{\phi(x_\phi)} \ge \frac{g(x)}{\phi(x)}, \qquad \textrm{for all } x\in\CI,
$$
but there is no maximal turning  point of $g/\psi$  in $\CI$. This case is associated with put option type payoffs.
\end{enumerate}

Having constructed Cases \ref{CALL}--\ref{PUT} by combining Cases \ref{GO}--\ref{STOP}, we construct the final two cases by combining Cases \ref{CALL}--\ref{PUT}.

\begin{enumerate}[{CASE }I]
\setcounter{enumi}{4}
\item \label{HILL}  $\Lop g$ is {positive} for some $x\in\mathcal{E}:=]\alpha,x_l[\,\cup\,]x_r,\beta[$ and  and negative for all $x \in [x_l,x_r]$ such that $g/\psi$  achieves a  maximal turning point for some $x_\psi\in[x_l,x_r]$ while $g/\phi$  achieves a  maximal turning point for some $x_\phi\in[x_\psi,x_r]$ with
$$
\frac{g(x_\psi)}{\psi(x_\psi)}\ge\frac{g(x)}{\psi(x)} \quad \textrm{and}\quad  \frac{g(x_\phi)}{\phi(x_\phi)} \ge \frac{g(x)}{\phi(x)} \quad\textrm{for all } x\in\CI.
$$  
This case is associated with butterfly option type payoffs.
\end{enumerate}  

\begin{enumerate}[{CASE }I]
\setcounter{enumi}{5}
\item \label{VALL} $\Lop g$ is {negative} for all $x\in\mathcal{E}:=]\alpha,x_l[\,\cup\,]x_r,\beta[$ and  positive for some $x \in [x_l,x_r]$ such that $g/\phi$  achieves a  stationary point for some $x_\phi\in]\alpha,x_l]$, 
 while $g/\psi$  achieves a  stationary point for some $x_\psi\in[x_r,\beta[$, with
\begin{gather} \label{ex-cross}
\lim_{x\downarrow\alpha} \frac{g(x)}{\psi(x)} \ge \frac{g(x_\psi)}{\psi(x_\psi)}\quad\textrm{and}\quad  \lim_{x\uparrow\beta} \frac{g(x)}{\phi(x)} \ge \frac{g(x_\phi)}{\phi(x_\phi)}.
\end{gather}
This case is associated with straddle option type payoffs.
\end{enumerate}

Observe that (\ref{e:R+phi})--(\ref{e:R-psi}) imply that Case \ref{HILL} occurs when we have $x_\psi\le x_\phi$, while Case \ref{VALL} occurs when $x_\phi\le x_\psi$.

%Figure \ref{cases} presents schematics for these six cases.
%\begin{figure}[ht]
% \begin{center} 
% \includegraphics{plots1.pdf}
%\caption{\label{cases} Schematics of Cases I--VI.}
%\end{center}
%\end{figure}

Case \ref{VALL} is important in that it represents situations when the stopping problem is one of the first exit time of the diffusion from an interval, rather than the the cases when one boundary is inaccessible and the problem is one of locating the first hitting time of a point, which characterise Cases \ref{GO}--\ref{HILL}.  Practically this means that the continuation region has two boundaries, one on the left hand side, which we shall denote by $a$, and the other on the right hand side, $b$, and the following conditions need to be satisfied at the two boundaries
\begin{align}
A \phi'(a )+B \psi'(a )\le &\spa g_-'(a ) &  A \phi'(b )+B \psi'(b )\le &\spa g_-'(b )\label{V1}\\
A \phi(a )+B \psi(a )=&\spa g(a ) &  A \phi(b )+B \psi(b )=&\spa g(b )\label{V2}\\
A \phi'(a )+B \psi'(a )\ge &\spa g_+'(a ) &  A \phi'(b )+B \psi'(b )\ge &\spa g_+'(b ).\label{V3}
\end{align}

We note that (\ref{V1})--(\ref{V3}) mean that the points $\{a,b\}$ define maximal turning  points of the function
\begin{gather*}
\frac{g(x)}{A\phi(x)+B\psi(x)},\quad \textrm{for } x\in\CI,
\end{gather*}
and that
\begin{gather*}
\frac{g(a)}{A\phi(a)+B\psi(a)}=\frac{g(b)}{A\phi(b)+B\psi(b)}=1.
\end{gather*}
These two observations are important in the martingale approach to solving stopping problems (such as in \cite{bl97},~\cite{bl00},~\cite{CI_HFTOS}).

In order to identify the locations of $\{a,b\}$, and hence the values for $A$ and $B$, observe that by using  (\ref{e:R+phi})--(\ref{e:R-psi}), (\ref{V1})--(\ref{V3}) can be rearranged into the following set of equations
\begin{gather}
\label{teqB-a} 
-\int_{]a ,\beta [} \frac{2 \phi(s)}{\sigma^2(s)\W(s)}\Lop g(ds) \le B  \le  -\int_{[a ,\beta [} \frac{2 \phi(s)}{\sigma^2(s)\W(s)}\Lop g(ds)\\
\label{teqB-b} 
-\int_{]b ,\beta [} \frac{2 \phi(s)}{\sigma^2(s)\W(s)}\Lop g(ds) \le B  \le -\int_{[b ,\beta [} \frac{2 \phi(s)}{\sigma^2(s)\W(s)}\Lop g(ds)\\
\label{teqA-a} 
-\int_{]\alpha ,a ]} \frac{2 \psi(s)}{\sigma^2(s)\W(s)}\Lop g(ds) \ge A  \ge -\int_{]\alpha ,a [} \frac{2 \psi(s)}{\sigma^2(s)\W(s)}\Lop g(ds)\\
\label{teqA-b} 
-\int_{]\alpha ,b ]} \frac{2 \psi(s)}{\sigma^2(s)\W(s)}\Lop g(ds)\ge A  \ge -\int_{]\alpha ,b [} \frac{2 \psi(s)}{\sigma^2(s)\W(s)}\Lop g(ds),
\end{gather}
These  are equivalent to the following system of equations
\begin{align}
\label{e:qphi} q_\phi^\op(a,b) \ge 0 \quad\textrm{and}\quad
q_\phi^\cl(a,b) \le 0 
\intertext{and}
\label{e:qpsi} q_\psi^\op(a,b) \ge 0 \quad\textrm{and}\quad
q_\psi^\cl(a,b) \le 0 ,
\end{align}
where
\begin{align}
\label{e:q_phi+} q_\phi^\cl(y,z):=&\spa  \int_{[y,z]} \frac{2 \phi(s)}{\sigma^2(s)\W(s)}\Lop g(ds) \\
\nn =&\spa \int_{[y,\beta [} \frac{2\phi(s)}{\sigma^2(s)\W(s)} \Lop g(ds) - \int_{]z,\beta [} \frac{2\phi(s)}{\sigma^2(s)\W(s)} \Lop g(ds),\\
\label{e:q_phi-} q_\phi^\op(y,z):=&\spa \int_{]y,z[} \frac{2\phi(s)}{\sigma^2(s)\W(s)} \Lop g(ds) \\
\nn =&\spa \int_{]y,\beta [} \frac{2\phi(s)}{\sigma^2(s)\W(s)} \Lop g(ds) - \int_{[z,\beta [} \frac{2\phi(s)}{\sigma^2(s)\W(s)} \Lop g(ds),
\intertext{and}
\label{e:q_psi-} q_\psi^\op(y,z):=&\spa \int_{]y,z[} \frac{2\psi(s)}{\sigma^2(s)\W(s)} \Lop g(ds)  \\
\nn =&\spa \int_{]\alpha ,z[} \frac{2\psi(s)}{\sigma^2(s)\W(s)} \Lop g(ds) - \int_{]\alpha ,y]} \frac{2\psi(s)}{\sigma^2(s)\W(s)} \Lop g(ds),\\
\label{e:q_psi+} q_\psi^\cl(y,z):=&\spa \int_{[y,z]} \frac{2 \psi(s)}{\sigma^2(s)\W(s)}\Lop g(ds)\\
\nn =&\spa \int_{]\alpha ,z]} \frac{2\psi(s)}{\sigma^2(s)\W(s)} \Lop g(ds) - \int_{]\alpha ,y[} \frac{2\psi(s)}{\sigma^2(s)\W(s)} \Lop g(ds).
\end{align}

We need to strengthen our assumptions on $g$ in order to prove the existence and uniqueness of $\{a,b\}$ appearing in  (\ref{e:qphi})--(\ref{e:qphi}), this is done, along with some explanation, in Lemma \ref{l:Vbound} in the Appendix.

We now solve the various  control problems described in Cases \ref{GO}--\ref{VALL} by constructing  explicit solutions of the variational inequalities  (\ref{e:hjb}) that satisfies the requirements of (\ref{vsol}), Definition~\ref{HJB-sense} and (\ref{w-domin}).  
\begin{thrm}\label{thrm:main_result}
Suppose that Assumptions~\ref{A1},\ref{A2},\ref{A3} and \ref{A4} hold.  We have the following solutions to the discretionary stopping problem we have formulated as Cases \ref{GO}--\ref{VALL}.\\\\
\textbf{ Case \ref{GO}. }Given any initial condition $x\in\CI$, then the value function $v$ is given by  $v(x)=0$ and $\Cr=\CI$.  In this case there is no admissible stopping strategy; the optimal stopping time is $\tau^*=\infty$.
\\\\
\textbf{ Case \ref{STOP}. } Given any initial condition $x\in\CI$, then the value function $v$ is given by $v(x)=g(x)$, $\Sr=\CI$ and the optimal stopping time is $\tau^*=0$.
\\\\
\textbf{ Case \ref{CALL}. } Given any initial condition $x\in\CI$, then   the value function $v$ is given by
\begin{align}\label{Csol}
 v(x) =&\spa \begin{cases}
     B \psi(x),& \textrm{if }  x\in \Cr = ]\alpha, x_\psi[,\\
        g(x), & \textrm{if }   x\in \Sr = [x_\psi, \beta[,\\
  \end{cases}
\end{align}
with  $B=g(x_\psi)/\psi(x_\psi)>0$. Furthermore, given any initial
condition $x\in\CI$, the stopping strategy $(\mathbb{S}^*_x, \tau^*)\in \, \mathcal{S}_x$, where
$\mathbb{S}^*_x$ is a weak solution to (\ref{SDE}) and 
\begin{align*}
  \tau^*\;=\;\inf \{t\ge0 \,|\  X_t \in \Sr \},
\end{align*}
is  optimal.
\\\\
\textbf{ Case \ref{PUT}. } Given any initial condition $x\in\CI$, then   the value function $v$ is given by
\begin{align}\label{Psol}
 v(x) =&\spa \begin{cases}
        g(x), & \textrm{if }   x\in \Sr = ]\alpha, x_\phi],\\
    A\phi(x) ,& \textrm{if }   x\in \Cr = ]x_\phi, \beta[,\\
  \end{cases}
\end{align}
with $A=g(x_\phi)/\phi(x_\phi)>0$. Furthermore, given any initial
condition $x\in\CI$, the stopping strategy $(\mathbb{S}^*_x, \tau^*)\in \, \mathcal{S}_x$, where
$\mathbb{S}^*_x$ is a weak solution to (\ref{SDE}) and 
\begin{align*}
  \tau^*\;=\;\inf \{t\ge0 \,|\  X_t \in \Sr \},
\end{align*}
is  optimal.
\\\\
\textbf{ Case \ref{HILL}. } Given any initial condition $x\in\CI$, then  $\Cr=]\alpha, x_\psi[\cup]x_\phi,\beta[$ and the value function $v$ is given by
\begin{align}\label{Hsol}
 v(x) =&\spa \begin{cases}
 B \psi(x),& \textrm{if }  x\in \Cr_1 = ]\alpha, x_\psi[\\
         g(x) ,& \textrm{if }  x\in \Sr = [x_\psi, x_\phi],\\
    A\phi(x) ,& \textrm{if }  x\in \Cr_2 = ]x_\phi,\beta[.
  \end{cases}
\end{align}
with  $B=g(x_\psi)/\psi(x_\psi)>0$ and $A=g(x_\phi)/\phi(x_\phi)>0$. Furthermore, given any initial
condition $x\in\CI$, the stopping strategy $(\mathbb{S}^*_x, \tau^*)\in \, \mathcal{S}_x$, where
$\mathbb{S}^*_x$ is a weak solution to (\ref{SDE}) and 
\begin{align*}
  \tau^*\;=\;\inf \{t\ge0 \,|\  X_t \in \Sr \},
\end{align*}
is  optimal.
\\\\
\textbf{ Case \ref{VALL}. } If  $-\Lop g$ is a positive, non-atomic measure on $(\mathcal{E}, \mathcal{B}(\mathcal{E}))$ and (\ref{ex-cross}) is true, then there exist a unique pair $(a,b)\in\mathcal{E}$ such that (\ref{e:qphi})--(\ref{e:qpsi}) are true.  In these circumstances, given any initial condition $x\in\CI$, then  $\Cr=]a,b[$ and the value function $v$ is given by
\begin{align}\label{Vsol}
 v(x) =&\spa \begin{cases}
         g(x) ,& \textrm{if }  x\in \Sr_1 =]\alpha,a],\\
    A\phi(x) + B \psi(x),& \textrm{if }  x\in \Cr = ]a,b[,\\
        g(x), & \textrm{if }   x\in \Sr_2 =[b,\beta[,\\
  \end{cases}
\end{align}
with 
\begin{align}
\label{e:A} A=&\spa \frac{g(b)\psi(a)-g(a)\psi(b)}{\phi(b)\psi(a)-\phi(a)\psi(b)},\\
\label{e:B} B=&\spa \frac{g(b)\phi(a)-g(a)\phi(b)}{\phi(a)\psi(b)-\phi(b)\psi(a)} .
\end{align}
Furthermore, given any initial
condition $x\in\CI$, the stopping strategy $(\mathbb{S}^*_x, \tau^*)\in \, \mathcal{S}_x$, where
$\mathbb{S}^*_x$ is a weak solution to (\ref{SDE}) and  
\begin{align*}
  \tau^*\;=\;\inf \{t\ge0 \,|\  X_t \in \Sr_1\,\cup\, \Sr_2 \},
\end{align*}
is  optimal.

\end{thrm}
We immediately note that for $v$ to be the difference of two convex functions, we will require continuous fit at the boundaries between the continuation/stopping regions.  This is satisfied by all of (\ref{Csol})--(\ref{Vsol}).  On this basis, the measure $\Lop v$ will be $(\phi, \psi)$-integrable given $\Lop \phi=\Lop \psi=0$ and since $\Lop g$ is $(\phi, \psi)$-integrable by Assumption \ref{A4}.
\\\\
\textbf{Proof of Case \ref{GO}. }   The stopping region cannot coincide with any interval in which $\Lop g$ is strictly positive, by (\ref{HJB1}), implying that $\Cr=\CI$ and (\ref{HJB3}) is satisfied.  To satisfy (\ref{vsol}) and  (\ref{w-domin}) we require that $A=B=0$, implying $v(x)=0$.  Equations  (\ref{e:R+phi})--(\ref{e:R-psi}) mean that $g/\psi$ is increasing (resp., $g/\phi$ is decreasing) for all $x\in\CI$, however, this and  (\ref{limg}) in Assumption \ref{A4} implies that $g(x)<0$ for all $x\in\CI$ and (\ref{HJB2}) is satisfied.
\\\\
\textbf{Proof of Case \ref{STOP}. }Since $\Lop g$ is negative for all $x\in\CI$,  Dynkin's formula (\ref{DYNKIN}) implies that it would be optimal to stop immediately and that $\Cr=\emptyset$ and (\ref{HJB1}) is true.  Since  $v(x)=g(x)$ for all $x\in\CI$, (\ref{HJB2})--(\ref{HJB3})    and  (\ref{w-domin}) are satisfied.

Note that equations (\ref{e:R+phi})--(\ref{e:R-psi}) mean that $g/\psi$ is decreasing (resp., $g/\phi$ is increasing) for all $x\in\CI$. This, with  (\ref{limg}) in Assumption \ref{A4} implies that $g(x)>0$ for all $x\in\CI$ and the value function is strictly positive.
\\
\\
\textbf{Proof of Case \ref{CALL}. } Observe that as a consequence of $\Lop g$ being {positive} for all $x\in]\alpha,x_l[$ and {negative} for all $x\in[x_r,\beta[$ and (\ref{e:R+psi})--(\ref{e:R-psi}) we will have that $x_\psi\in[x_r,\beta[$.  In addition, $g(x_\psi)/\psi(x_\psi)>0$  because $g/\psi$ is decreasing in $]x_\psi,\beta[$ and $\lim_{x\uparrow\beta} g/\psi=0$, the implication being that $B>0$.  Also, (\ref{Csol}) satisfies (\ref{HJB2}) because 
$$
\frac{g(x)}{\psi(x)} \le \frac{g(x_\psi)}{\psi(x_\psi)}=B, \quad \textrm{for all }
  x\in\CI.
$$
For $v$ given by (\ref{Csol}), (\ref{HJB3}) is true and, similarly, (\ref{HJB1}) is true since $x_\psi\in[x_r,\beta[$ and $\Lop g<0$ in this interval. Finally,  (\ref{w-domin}) holds since $\sup_{x\in\Cr} v(x) \le g(x_\psi)<1+g(x_\psi)$.
\\\\
%%%%%%%%%%%%%%%%%%%%%%%%%%%%%%%%%%%%%%%%%%%%%%%%%%%%%%%%%%%%%%%%%%%%%%%%%%%%%%%%%%%%%%%%%%%%%%%%%%%%%%%%%%%%%%%%%%%
%%%%%%%%%%%%%%%%%%%%%%%%%%%%%%%%%%%%%%%%%%%%%%%%%%%%%%%%%%%%%%%%%%%%%%%%%%%%%%%%%%%%%%%%%%%%%%%%%%%%%%%%%%%%%%%%%%%
\textbf{Proof of Case \ref{PUT}. }This is symmetric to Case \ref{CALL}.  Observe that as a consequence of $\Lop g$ being {negative} for all $x\in]\alpha,x_l[$ and {positive} for all $x\in[x_l,\beta[$ and (\ref{e:R+phi})--(\ref{e:R-phi}) we will have that $x_\phi\in]\alpha,x_l]$ implying, as with Case \ref{CALL}, $A= g(x_\phi)/\phi(x_\phi)>0$.  Since
$$
\frac{g(x)}{\phi(x)} \le \frac{g(x_\phi)}{\phi(x_\phi)}=A, \qquad \textrm{for all } x\in\CI,
$$
(\ref{Psol}) satisfies (\ref{HJB2})--(\ref{HJB3}), (\ref{HJB1}) is true since $x_\psi\in\in[x_l,\beta[$ and $\Lop g<0$ in this interval. Finally,  (\ref{w-domin}) holds since $\sup_{x\in\Cr} v(x) \le g(x_\phi)<1+g(x_\phi)$.
\\\\%%%%%%%%%%%%%%%%%%%%%%%%%%%%%%%%%%%%%%%%%%%%%%%%%%%%%%%%%%%%%%%%%%%%%%%%%%%%%%%%%%%%%%%%%%%%%%%%%%%%%%%%%%%%%%%%%%%
%%%%%%%%%%%%%%%%%%%%%%%%%%%%%%%%%%%%%%%%%%%%%%%%%%%%%%%%%%%%%%%%%%%%%%%%%%%%%%%%%%%%%%%%%%%%%%%%%%%%%%%%%%%%%%%%%%%
\textbf{Proof of Case \ref{HILL}. }We can regard Case \ref{HILL} as being composed of (moving from $\alpha$ to $\beta$) Case \ref{CALL}, Case \ref{STOP}, and then Case \ref{PUT}.  The proof of this case is a combination of the proofs of these more elementary cases.  
\\\\%%%%%%%%%%%%%%%%%%%%%%%%%%%%%%%%%%%%%%%%%%%%%%%%%%%%%%%%%%%%%%%%%%%%%%%%%%%%%%%%%%%%%%%%%%%%%%%%%%%%%%%%%%%%%%%%%%%
%%%%%%%%%%%%%%%%%%%%%%%%%%%%%%%%%%%%%%%%%%%%%%%%%%%%%%%%%%%%%%%%%%%%%%%%%%%%%%%%%%%%%%%%%%%%%%%%%%%%%%%%%%%%%%%%%%%
\textbf{Proof of Case \ref{VALL}. }We begin by noting that   Lemma \ref{l:Vbound} proves the  existence of a unique pair $(a,b)\in\mathcal{E}$ such that (\ref{e:qphi})--(\ref{e:qpsi}) are true.  To see that (\ref{Vsol}) satisfies (\ref{HJB2}), recall that (\ref{e:qphi})--(\ref{e:qpsi}) imply that  the points $\{a,b\}$ define maximal turning points of the function
\begin{gather*}
\frac{g(x)}{A\phi(x)+B\psi(x)},\quad \textrm{for } x\in\CI,
\end{gather*}and so
\begin{gather*}
A\phi(x) + B \psi(x) \ge g(x),\quad \textrm{for all } x\in\,\Cr.
\end{gather*}

Also, for $v$ given by (\ref{Vsol}), (\ref{HJB3}) is true and, similarly, (\ref{HJB1}) is true since $\{a,b\}\in \mathcal{E}$,  while  (\ref{w-domin}) holds since $$\sup_{x\in\Cr} v(x) = \max\{g(a),g(b)\}<1+\max\{g(a),g(b)\}.$$

To see that $B>0$ observe that, from (\ref{teqB-a})
\begin{align*}
B \ge& -\int_{]a ,\beta [} \frac{2 \phi(s)}{\sigma^2(s)\W(s)}\Lop g(ds)\\
=& -\int_{]a ,x_\phi ]} \frac{2 \phi(s)}{\sigma^2(s)\W(s)}\Lop g(ds)  -\int_{[x_\phi ,\beta [} \frac{2 \phi(s)}{\sigma^2(s)\W(s)}\Lop g(ds)\\
\ge& 0,
\end{align*}
with the final inequality being a consequence of $\Lop g$ being {negative} for all $x\in]\alpha,x_\phi]$ and 
$$
-\int_{[x_\phi ,\beta [} \frac{2 \phi(s)}{\sigma^2(s)\W(s)}\Lop g(ds) \ge 0
$$
since $x_\phi$ is a maximum turning point of $g/\phi$.  Similar arguments show that $A>0$.\\\\
%%%%%%%%%%%%%%%%%%%%%%%%%%%%%%%%%%%%%%%%%%%%%%%%%%%%%%%%%%%%%%%%%%%%%%%%%%%%%%%%%%%%%%%%%%%%%%%%%%%%%%%%%%%%%%%%%%%
%%%%%%%%%%%%%%%%%%%%%%%%%%%%%%%%%%%%%%%%%%%%%%%%%%%%%%%%%%%%%%%%%%%%%%%%%%%%%%%%%%%%%%%%%%%%%%%%%%%%%%%%%%%%%%%%%%%
\mbox{}\hfill$\Box$

\begin{rem}\label{rem0}
 The justification for the `pasting' of intervals in Case \ref{HILL} comes from Lamberton and Zervos \cite[i.e. Theorem 12]{LZ_OSODD}, where the stopping problem is solved for the case of absorbing boundaries with the payoff at the boundaries being positive, with the result that  the absorbing boundaries are in the stopping region.  In Case \ref{HILL},  the value function at the boundaries of the sub-intervals ($x_\psi$ and $x_\phi$) identifies with the payoff and we can regard the open interval $\CI$ being made up of three separate stopping problems on $]\alpha,x_\psi]$, $[x_\psi,x_\phi]$ and $[x_\phi,\beta[$ with the diffusion being `absorbed' at $x_\psi$ and $x_\phi$.
\end{rem}

\begin{rem}\label{rem1}
Notice that in Cases \ref{CALL}--\ref{HILL}, since it is not a requirement that $\Lop g<0$ in the waiting region, it is necessary and sufficient  that there are maximal turning points at $x_\psi$ (resp., $x_\phi$) such that
\begin{gather*}
 \frac{g(x)}{\psi(x)} \le \frac{g(x_\psi)}{\psi(x_\psi)},\quad \textrm{for all } x\in]\alpha,x_\psi]\\
\left(\textrm{resp., } \frac{g(x)}{\phi(x)} \le \frac{g(x_\phi)}{\phi(x_\phi)}\quad \textrm{for all } x\in]x_\phi,\beta]\right).
\end{gather*} 
\end{rem}
\begin{rem}\label{rem2}
 The existence of the stationary points $x_\phi$ and $x_\psi$ is sufficient, not necessary in Case~\ref{VALL}.  There will be cases when $\Lop g >0$ for some interval, but this will not lead to a stationary point of either of the functions $g/\phi$ or $g/\psi$.
\end{rem}

\section{Examples}\label{ex}
We finish with some concrete examples to  develop the ideas presented in Section \ref{soln}. In all the examples we shall consider the   diffusion, $X$, is given by a geometric Brownian motion with drift $b=0$ and volatility $\sigma=0.2$ and there is a constant discount rate of $r=0.01$.  In this case, as presented in Section \ref{implications}, we have that
\begin{gather*}
 \phi(x) = x^{\hf-\sqrt{\frac{3}{4}}}\quad\textrm{and}\quad\phi(x) = x^{\hf+\sqrt{\frac{3}{4}}}.
\end{gather*}

In the first example, we consider a payoff function of the form
\begin{align*}
 g(x) =&\spa \begin{cases}
 c,& \textrm{if }  x< 2\\
 x+c-2,& \textrm{if }  2 \le x, 
  \end{cases}, \quad\textrm{with }c\ge-2,
\end{align*}
which satisfies Assumption \ref{A4}. In this case
$$
\Lop g(dx) = 0.08 \delta_2 - 0.01 g(x),
$$
where $\delta_y$ is the Dirac probability measure, assigning mass 1 at the point $\{y\}$.  

This means that if $c\ge 8$, $-\Lop g$ is a positive measure for all $x \in \CI$  and we have the conditions of Case \ref{STOP}.  If $c\in[-2,0]$ we have the conditions for Case \ref{CALL}.  In the event that $c\in\;]0,8[\,$, we know we will be waiting at $x=2$ since $\Lop g\{2\}>0$ and while we do not have the specific conditions for Case \ref{VALL}, in fact there is no stationary point of $g/\psi$ for $c>1.5$, we can infer that the value function is going to be of the form (\ref{Vsol}), since in the cases when we do have a stationary point of $g/\psi$,
$$
\lim_{x\downarrow\alpha} \frac{g(x)}{\psi(x)} > \frac{g(x_\psi)}{\psi(x_\psi)}.
$$

When $c\in\;]0,8[\,$, the boundaries of the stopping/continuation regions, $\{a,b\}$, can be found either by solving the the smooth fit problem
\begin{align*}
 A\phi(a)+B\psi(a) &=\spa c,& A\phi(b)+B\psi(b) &=\spa b+c-2\\
A\phi'(a)+B\psi'(a) &=\spa 0,& A\phi'(b)+B\psi'(b) &=\spa b,
\end{align*}
or by finding the unique $\{a,b\}$ such that $a=a_\phi=a_\psi$ and $b=b_\phi=b_\psi$ where $a_\phi$ and $b_\phi$ solve
\begin{align}
 c \int_{a_\phi}^{b_\phi} \frac{2 \phi(s)}{\sigma^2(s)\W(s)} ds  +   \int_2^{b_\phi} \frac{2 \phi(s)}{\sigma^2(s)\W(s)}\, (s-2)\, ds = &\ 8\frac{2 \phi(2)}{\sigma^2(2)\W(2)},\label{S1}
\intertext{while $a_\psi$ and $b_\psi$ solve}
c \int_{a_\psi}^{b_\psi} \frac{2 \psi(s)}{\sigma^2(s)\W(s)} ds  +   \int_2^{b_\psi} \frac{2 \psi(s)}{\sigma^2(s)\W(s)}\,(s-2)\, ds =&\ 8\frac{2 \psi(2)}{\sigma^2(2)\W(2)}.\label{S2}
\end{align}
One benefit of using (\ref{S1})--(\ref{S2}) over `smooth fit', is that it is immediately obvious that there is no solution   with $A,B\ge0$ when $c\ge 8$, which is not obvious \emph{a priori} from the four smooth fit equations. To appreciate how this method has advantages over the `martingale approach', see \cite[Example 5]{LZ_OSODD}.

Setting $c=2$, (\ref{S1})--(\ref{S2}) can be expressed as
\begin{align*}
a_\phi=&\spa\sqrt[m-1]{(m-1)\left(2^{m-1}\left(\frac{1}{m-1}-\frac{1}{m}-2\right) +\frac{b_\phi^m}{2m} \right)}=:\;l_\phi(b_\phi),
\intertext{and,}
a_\psi=&\spa\sqrt[n-1]{(n-1)\left(2^{n-1}\left(\frac{1}{n-1}-\frac{1}{n}-2\right)+\frac{b_\psi^n}{2n}  \right)}=:\;l_\psi(b_\psi),
\end{align*}
with $m= \hf-\sqrt{\frac{3}{4}}$ and $n= \hf+\sqrt{\frac{3}{4}}\,$.  By finding the crossing of $l_\phi$ and $l_\psi$, we  deduce that, in this case, the value function is given by
\begin{align*}
 v(x) =&\spa \begin{cases}
         2 ,& \textrm{if }  x\in \Sr_1 =]\alpha,0.9350],\\
    1.5439\,x^{\hf-\sqrt{\frac{3}{4}}} + 0.4578\,x^{\hf+\sqrt{\frac{3}{4}}},& \textrm{if }  x\in \Cr = ]0.9350,5.2335[,\\
        x, & \textrm{if }   x\in \Sr_2 =[5.2335,\beta[.\\
  \end{cases}
\end{align*}

Our second example involves two functions that do not satisfy Assumption \ref{A4}, but never the less demonstrate the usefullness of considering complex stopping problems in terms of Cases \ref{GO}--\ref{VALL}.  Consider two `staircase' type payoffs, as discussed in Bronstein, et al. \cite{BHPZ}, 
\begin{align*}
 g_1(x) =&\spa \begin{cases}
 0,& \textrm{if }  x< 2,\\
 1,& \textrm{if }  2 \le x< 4,\\
 4,& \textrm{if }  4 \le x< 6,\\
 9,& \textrm{if }  6 \le x< 8,\\
 16,& \textrm{if }  8 \le x< 10,\\
 25,& \textrm{if }  10 \le x,
  \end{cases}
&
 g_2(x) =&\spa \begin{cases}
 0,& \textrm{if }  x< 2,\\
 2,& \textrm{if }  2 \le x< 4,\\
 4,& \textrm{if }  4 \le x< 6,\\
 6,& \textrm{if }  6 \le x< 8,\\
 8,& \textrm{if }  8 \le x< 10,\\
 10,& \textrm{if }  10 \le x.
  \end{cases}
\end{align*}

Since these functions are not continuous, they do not satisfy the conditions of Assumption \ref{A4} apart from (\ref{limg}).  However, Lamberton and Zervos make the observation that any points at which the payoff function is discontinuous will be part of the continuation region, and so it is possible to construct a value function that conforms with Definition \ref{HJB-sense}.  The analysis presented in this paper relies on Assumption \ref{A4} in the widespread use of expressions such as (\ref{e:R+phi})--(\ref{e:R-psi}), which are applied, for example, in the proof of  Lemma \ref{l:Vbound}.  However the basic intuition of considering the sign of $\Lop g$ and stationary points of the functions $g/\phi$ and $g/\phi$ can still be helpful when considering these `staircase' payoff functions.

There are turning points of $g_{(1,2)}/\psi$ at $2,4,6,8$ and $10$,  and we note that
$$
0.3880=\frac{g_1}{\psi}(2) < \frac{g_1}{\psi}(4) < \frac{g_1}{\psi}(6) < \frac{g_1}{\psi}(8) < \frac{g_1}{\psi}({10})=1.1.0763,
$$
while
$$
0.7759=\frac{g_2}{\psi}(2) > \frac{g_2}{\psi}(4) > \frac{g_2}{\psi}(6) > \frac{g_2}{\psi}(8) > \frac{g_2}{\psi}({10})=0.4305.
$$
%These situations are described in Figure \ref{step}.
%\begin{figure}[ht]
% \begin{center} 
% \includegraphics{step.pdf}
%\caption{\label{step} Schematics of plots for $g_1$ and $g_2$.}
%\end{center}
%\end{figure}

These sequences mean that the solution to the two problems are very different.  With $g_1$ there is a global maximum turning point at $x=10$, and we have Case \ref{CALL} with
\begin{align*}
 v_1(x) =&\spa \begin{cases}
     1.1.0763 \,x^{\hf+\sqrt{\frac{3}{4}}} ,& \textrm{if }  x\in \Cr = ]\alpha, 10[,\\
        25, & \textrm{if }   x\in \Sr = [10, \beta[,\\
  \end{cases}
\end{align*}
and (\ref{HJB1})--(\ref{w-domin}) are satisfied.  

Now, with $g_2$ we have  the situation of a series of sub-intervals, as described in Remark \ref{rem0}.  We have Case \ref{CALL} in the interval $]0,2]$,  Case \ref{VALL} for the intervals $[2,4],[4,6],[6,8],[8,10]$ and Case \ref{STOP} for $[10,\beta[$.  To develop our understanding of the four versions of Case \ref{VALL}, define each jump location as $j=4,6,8,10$.  The right-hand boundary of the four intervals must be continuous fit at $j$, while, employing (\ref{V1})--(\ref{V3}), the left hand boundary will satisfy smooth fit (see also \cite[Lemma 4]{BHPZ}) if
$$
\frac{\psi'(j-2)}{\phi'(j-2)}\le \frac{j\psi(j-2)-(j-2)\psi(j)}{j\phi(j-2)-(j-2)\phi(j)}.
$$
If this condition is not satisfied, we will also have only continuous fit at the left hand boundary, $(j-2)$. In the case under consideration, it can be deduced that $\Sr = \big\{\{2\},\{4\},\{6\},\{8\},\{10\}\big\}$   and it is easy to establish that 
\begin{align*}
 v_2(x) =&\spa \begin{cases}
     0.7759 \,x^{\hf+\sqrt{\frac{3}{4}}} ,& \textrm{if }  x\in\,  ]\alpha, 2[,\\
        2, & \textrm{if }   x=2,\\
     0.8263 \,x^{\hf-\sqrt{\frac{3}{4}}} \,+\, 0.5272 \,x^{\hf+\sqrt{\frac{3}{4}}} ,& \textrm{if }  x\in\, ]2, 4[,\\
        4, & \textrm{if }   x=4,\\
     1.8162 \,x^{\hf-\sqrt{\frac{3}{4}}} \,+\, 0.4375 \,x^{\hf+\sqrt{\frac{3}{4}}} ,& \textrm{if }  x\in\, ]4, 6[,\\
        6, & \textrm{if }   x=6,\\
     2.9443 \,x^{\hf-\sqrt{\frac{3}{4}}} \,+\, 0.3868 \,x^{\hf+\sqrt{\frac{3}{4}}} ,& \textrm{if }  x\in\, ]6, 8[,\\
        8, & \textrm{if }   x=8,\\
     4.1899 \,x^{\hf-\sqrt{\frac{3}{4}}} \,+\, 0.3528 \,x^{\hf+\sqrt{\frac{3}{4}}} ,& \textrm{if }  x\in\, ]8, 10[,\\
        10, & \textrm{if }   x\ge 10,
  \end{cases}
\end{align*}
and (\ref{HJB1})--(\ref{w-domin}) are satisfied. 

\appendix\section{Finding $\{a,b\}$ appearing in (\ref{e:qphi})--(\ref{e:qpsi})}

Observe that the construction of Case \ref{VALL} means that we will have a maximum turning point of $g/\phi$ at $x_\phi$ and a maximum turning point of $g/\psi$ at $x_\psi$ with 
\begin{gather}
 \alpha<x_\phi<x_l<x_r<x_\psi<\beta.
\end{gather}
Also, given $g/\phi$ is increasing for $x\downarrow\alpha$ while $g/\psi$ is decreasing for $x\uparrow\beta$, combined with (\ref{e:R-psi}) and (\ref{e:q_phi-}), and (\ref{e:R-phi}) with (\ref{e:q_psi-}), means that
\begin{gather}\label{c:q-op-ok}
q_\psi^\op(\alpha , \beta) < 0,\quad\textrm{and}\quad 
q_\phi^\op(\alpha , \beta) < 0,
\intertext{and}
\label{c:xlxr} q_\phi^\cl(x_l,x_r) > 0,\quad\textrm{and}\quad q_\psi^\cl(x_l,x_r)> 0.
\end{gather}
Now, for any $u\in\,]\alpha,x_l[$ and $v\in\,]x_r,\beta[$,  by decreasing $u$ or increasing $v$, $q_\phi^\cl(u,v)$ and $q_\phi^\cl(u,v)$ will decrease.  In addition, because we have maxima of $g/\phi$ and $g/\psi$,
\begin{gather}\label{l_phi_max}
q_\phi^\op(x_\phi , \beta) \ge 0,\quad\textrm{and}\quad 
q_\phi^\cl(x_\phi , \beta) \le 0,\\
\label{l_psi_min}
q_\psi^\op(\alpha , x_\psi) \ge 0,\quad\textrm{and}\quad 
q_\psi^\cl(\alpha , x_\psi) \le 0,
\end{gather} 
and we can see that the maximum value that $a$ could possibly take on, is $x_\phi$, while the minimum value $b$ could take on is $x_\psi$.  Similarly, there is a minimum value, $\overleftarrow{u}$, such that
\begin{gather}\label{l_phi_min}
q_\phi^\op(\overleftarrow{u}, \overleftarrow{v}) \ge 0,\quad\textrm{and}\quad 
q_\phi^\cl(\overleftarrow{u}, \overleftarrow{v}) \le 0.
\intertext{and a maximum value, $\overrightarrow{u}$,such that}\label{l_psi_max}
q_\psi^\op(\overrightarrow{u} , \overrightarrow{v}) \ge 0,\quad\textrm{and}\quad 
q_\psi^\cl(\overrightarrow{u} , \overrightarrow{v}) \le 0.
\end{gather}

As a consequence of (\ref{l_phi_max})--(\ref{l_psi_max}), there exist mappings  $l_\phi$ and $l_\psi$ such that 
\begin{gather}
\label{e:lphi} l_\phi:\ ]\overleftarrow{u},x_\phi] \,\mapsto\,[\overleftarrow{v},\beta[,\\
\label{e:lpsi} l_\psi:\ ]\alpha,\overrightarrow{u}] \,\mapsto\,[x_\psi,\overrightarrow{v}[,
\intertext{and}
\nn q_\phi^\op(y,l_\phi(y))\ge 0,\quad q_\phi^\cl(y,l_\phi(y))\le 0,\quad \textrm{ with }\quad y<l_\phi(y)\\
\nn q_\psi^\op(z,l_\psi(z)) \ge 0, \quad 
q_\psi^\cl(z,l_\psi(z)) \le 0,\quad \textrm{ with }\quad z<l_\psi(z).
\end{gather}
Finding a unique pair ($a,b$) is equivalent to  showing that there exists a unique crossing point of $l_\phi$ and $l_\psi$, where $y=z=a$ and $l_\phi(y) =l_\psi(z) =b$.

\begin{figure}[h]
 \begin{center} 
 \includegraphics{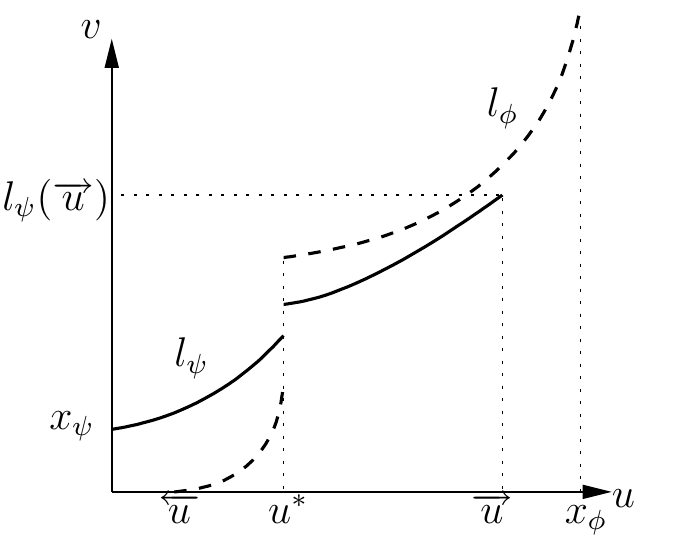}
\caption{\label{amb1}The effect of an atom in $\Lop g$ at $u^*$.}
\end{center}
\end{figure}

The first problem with proving this is that if the measure $\Lop g$ is \emph{atomic}, meaning that $\int  \Lop g (ds)$ could be discontinuous.  In this case, there would be discontinuities in $l_\phi$ and $l_\psi$ and we could have the situation described in Figure \ref{amb1}, and it is ambiguous where the crossing point of $l_\phi$ and $l_\psi$ is (in respect to the second parameter).   These issues do not occur if $\Lop g$ is non-atomic in $\mathcal{E}$.

\begin{figure}[h]
 \begin{center} 
 \includegraphics{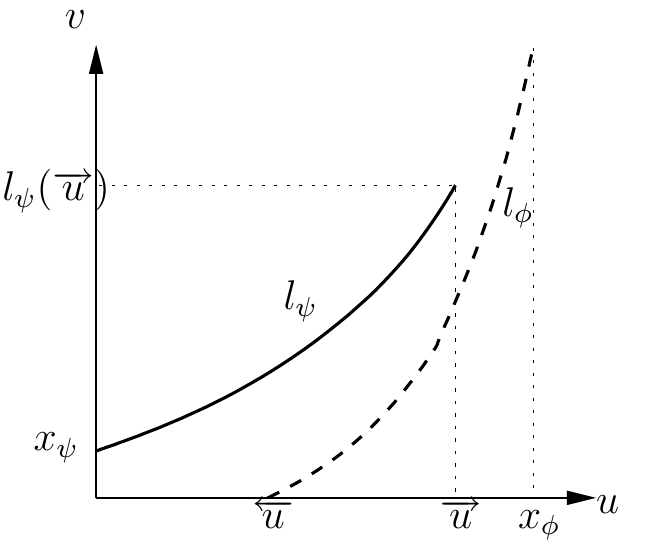}
\caption{\label{amb2} The effect of  being too far from $x_\phi$.}
\end{center}
\end{figure}

The second problem occurs if there is a significant gap between $x_\phi$ and $\overrightarrow{u_\psi}$ resulting in the situation described in Figure \ref{amb2}, there is no actual crossing point.   These issues do not occur if the boundaries $\alpha$ and $\beta$ are natural, as assumed in Lempa \cite[i.e. Lemma 2.1]{L_OSTS}.

\newpage\begin{lem}\label{l:Vbound} Suppose that
Suppose that Assumptions~\ref{A1},\ref{A2},\ref{A3} and \ref{A4} hold.  In addition 
\begin{enumerate}
 \item $\Lop g$ is  non-atomic on $\mathcal{E}\subset\CI$, as defined above, and $-\Lop g$ is a positive measure on $(\mathcal{E}, \mathcal{B}(\mathcal{E}))$.  
\item The following two inequalities hold:
\begin{gather}
\label{ex-cross2} \lim_{x\downarrow\alpha} \frac{g}{\psi}(x) \ge \frac{g}{\psi}(x_\psi)\quad{and}\quad  \lim_{x\uparrow\beta} \frac{g}{\phi}(x) \ge \frac{g}{\phi}(x_\phi)
\end{gather}
\end{enumerate}
then there exist a unique pair $(a,b)\in\mathcal{E}$ such that (\ref{e:qphi})--(\ref{e:qpsi}) are true.

\end{lem}
\textbf{Proof: } Our aim is to show that there exist unique mappings (\ref{e:lphi}--(\ref{e:lpsi}) such that 
\begin{gather*}
q_\phi(u_\phi,l_\phi(u_\phi)):= \int_{u_\phi}^{l_\phi(u_\phi)} \frac{2 \phi(s)}{\sigma^2(s)\W(s)}\Lop g(s)ds=0\\
q_\psi(u_\psi,l_\psi(u_\psi)):=\int_{u_\psi}^{l_\psi(u_\psi)} \frac{2 \psi(s)}{\sigma^2(s)\W(s)}\Lop g(s)ds=0\\
\textrm{with}\quad u_\phi<l_\phi(u)\textrm{ and }u_\psi< l_\psi(u)
\end{gather*}
and there is a unique crossing point of $l_\phi$ and $l_\psi$, where $a=u_\phi=u_\psi$ and $b=l_\psi(u)=l_\phi(u)$, $\Lop g(a)<0$.  In order to do this, let us first assume that such a $l_\phi$ and $l_\psi$ exist, and we shall first prove uniqueness.

We begin by noting that since $ x_\phi<x_l<x_r<x_\psi$, the points $(a,b)$, if they exist, will be in $\mathcal{E}=\CI/[x_l,x_r]$.

Now, observe that since
\begin{align*}
% q_\psi(u_\psi, l_\psi(u_\psi))=&\spa 0\\
dq_\psi(u_\psi, l_\psi(u_\psi))=&\spa \pd{q_\psi(u_\psi, l_\psi(u_\psi))}{l_\psi(u_\psi)}dl_\psi(u_\psi) + \pd{q_\psi(u_\psi, l_\psi(u_\psi))}{u_\psi}du_\psi = 0
\end{align*}
then
\begin{align*}
l_\psi'(u_\psi) %=&\spa  -\frac{\pd{q_\psi(u_\psi, l_\psi(u_\psi))}{u_\psi}}{\pd{q_\psi(u_\psi, l_\psi(u_\psi))}{l_\psi(u_\psi)}}\\
=&\spa -\frac{-\frac{2 \psi(u_\psi)}{\sigma^2(u_\psi)\W(u_\psi)}\Lop g(u_\psi)}{\frac{2 \psi(l_\psi(u_\psi))}{\sigma^2(l_\psi(u_\psi))\W(l_\psi(u_\psi))}\Lop g(l_\psi(u_\psi))}\\
>&\spa 0\qquad\textrm{for }u_\psi, l_\psi(u_\psi) \in \mathcal{E},
\intertext{and similarly from $q_\phi$, }
%q_\phi(u_\phi, l_\phi(u_\phi))=&\spa 0\\
% dq_\phi(u_\phi, l_\phi(u_\phi))=&\spa \pd{q_\phi(u_\phi, l_\phi(u_\phi))}{l_\phi(u_\phi)}dl_\phi(u_\phi) + \pd{q_\phi(u_\phi, l_\phi(u_\phi))}{u_\phi}du_\phi = 0\\
%\Rightarrow\qquad 
l_\phi'(u_\phi) %=&\spa  -\frac{\pd{q_\psi(u_\phi, l_\phi(u_\phi))}{u_\phi}}{\pd{q_\psi(u_\phi, l_\phi(u_\phi))}{l_\phi(u_\phi)}}\\
=&\spa -\frac{-\frac{2 \phi(u_\phi)}{\sigma^2(u_\phi)\W(u_\phi)}\Lop g(u_\phi)}{\frac{2 \phi(l_\phi(u_\phi))}{\sigma^2(l_\phi(u_\phi))\W(l_\phi(u_\phi))}\Lop g(l_\phi(u_\phi))}\\
>&\spa 0\qquad\textrm{for }u_\phi, l_\phi(u_\psi) \in \mathcal{E},
\end{align*}
with the inequalities being a consequence of the fact that $\Lop g(x)<0$ in $\mathcal{E}$.
Comparing $l_\psi'$ and $l_\phi'$  we can see
\begin{align*}
\frac{l_\phi'(x)}{l_\psi'(x)} %=&\spa\frac{\frac{\frac{2 \phi(x)}{\sigma^2(x)\W(x)}\Lop g(dx)}{\frac{2 \phi(l_\phi(x))}{\sigma^2(l_\phi(x))\W(l_\phi(x))}\Lop g(dl_\phi(x))}}{\frac{\frac{2 \psi(x)}{\sigma^2(x)\W(x)}\Lop g(dx)}{\frac{2 \psi(l_\psi(x))}{\sigma^2(l_\psi(x))\W(l_\psi(x))}\Lop g(dl_\psi(x))}}\\
%=&\spa \frac{\frac{2 \phi(x)}{\sigma^2(x)\W(x)}\Lop g(dx)}{\frac{2 \phi(l_\phi(x))}{\sigma^2(l_\phi(x))\W(l_\phi(x))}\Lop g(dl_\phi(x))} \frac{\frac{2 \psi(l_\psi(x))}{\sigma^2(l_\psi(x))\W(l_\psi(x))}\Lop g(dl_\psi(x))}{\frac{2 \psi(x)}{\sigma^2(x)\W(x)}\Lop g(dx)}\\
%=&\spa \frac{ \phi(x)}{\frac{2 \phi(l_\phi(x))}{} \frac{\frac{2 \psi(l_\psi(x))}{}}{\psi(x)}\\
=&\spa \frac{ \phi(x)}{ \phi(l_\phi(x))}\frac{ \psi(l_\psi(x))}{ \psi(x)}\frac{\Lop g(dl_\psi(x))}{\Lop g(dl_\phi(x))}\frac{\sigma^2(l_\phi(x))\W(l_\phi(x))}{\sigma^2(l_\psi(x))\W(l_\psi(x))},%\quad \textrm{for all } x \textrm{ such that } l_\phi(x) \textrm{ and } l_\psi(x) \textrm{ exist.} %\in\,]\alpha_\SI,a^{\rightarrow}_\psi[\, \cap \,]a^{\leftarrow}_\phi,\beta_\SI[ .
\end{align*}
Note that if there is a crossing  of $l_\phi$ and $l_\psi$ at $a=u_\phi=u_\psi$ and $b=l_\psi(u)=l_\phi(u)$ with $a<b$,
\begin{gather*}
 \frac{l_\phi'(a)}{l_\psi'(a)}=\frac{ \phi(a)}{ \phi(b)}\frac{ \psi(b)}{ \psi(a)}\frac{\Lop g(b)}{\Lop g(b)}\frac{\sigma^2(b)\W(b)}{\sigma^2(b)\W(b)}=\frac{ \phi(a)}{ \phi(b)}\frac{ \psi(b)}{ \psi(a)}>1
\end{gather*}
since $\phi$ is a decreasing function while $\psi$ is increasing.  This means that at a crossing $l_\phi'(x)>l_\psi'(x) $ and so if the crossing exists  $l_\phi$ crosses from below $l_\psi$. If this is the only way a crossing can occur, there can only be one such crossing.

We now show that a crossing exists.  Noting that, if they cross, then $l_\phi$ crosses $l_\psi$ from below then the crossing point exists  if 
\begin{gather*}
 l_\phi(\overleftarrow{u})\le  l_\psi(\overleftarrow{u})\quad\textrm{and}\quad l_\phi(\overrightarrow{u})\ge  l_\psi(\overrightarrow{u}).
\end{gather*}
Let us consider the second inequality.  We have that $l_\phi(x_\phi)=\beta$, and so if $l^{-1}_\psi(\beta)\in]\alpha,x_\phi]$ then this will hold.  Now, given the definition of $\mathcal{E}$, $l^{-1}_\psi(\beta)\in]\alpha,x_\phi]$ will be true if and only if
$$
\int_{x_\phi}^\beta \Psi(s)\Lop g(s) ds \le 0.
$$
Recall that (\ref{Rm-A}), that
\begin{gather*}
 -g(x) = \phi(x) \int_\alpha^x \Psi(s) \Lop g(s) ds + \psi(x)\int_x^\beta \Phi(s) \Lop g(s) ds,
\end{gather*}
and  so
$$
-\frac{g}{\phi}(\beta) = \int_\alpha^\beta \Psi(s) \Lop g(s) ds.
$$
Also, since $x_\phi$ is a turning point of $g/\phi$ we can see that
$$
-\frac{g}{\phi}(x_\phi) = \int_\alpha^{x_\phi} \Psi(s) \Lop g(s) ds.
$$
So
\begin{align*}
\int_{x_\phi}^\beta \Psi(s)\Lop g(s) ds =& \int_\alpha^\beta \Psi(s) \Lop g(s) ds- \int_\alpha^{x_\phi} \Psi(s) \Lop g(s) ds\\
=& \frac{g}{\phi}(x_\phi) -\frac{g}{\phi}(\beta)\\
\le& 0
\end{align*}
with the final inequality being a consequence of the second inequality in (\ref{ex-cross2}).  

The proof of $l_\phi(\overleftarrow{u})\le  l_\psi(\overleftarrow{u})$ follows symmetric arguments.
\mbox{}\hfill$\Box$

The condition (\ref{ex-cross2}) is sufficient, if it does hold it does not mean that the crossing point does not exist.

\begin{rem}\label{r:l}
In the cases where $\phi$ and $\psi$ are given by functions such as confluent hypogeometric functions or cylinder parabolic functions, solving (\ref{e:qphi})--(\ref{e:qpsi}) may be difficult.  Noting that solving (\ref{e:qphi})--(\ref{e:qpsi}) is equivalent to solving $l_\phi(x)-l_\psi(x) = 0$ and we have expressions for $l_\psi', l_\psi'$, the approach taken here may facilitate the solution to the problem.
\end{rem}

\mbox{}\hspace{-6mm}{\Large\textbf{Acknowledgements}}\mbox{}\vspace{3mm}

\mbox{}\hspace{-6mm}I would like to thank Professor Mihail Zervos for his care and inspiration.  In addition I would like to thank, amongst others,  Professor Damien Lamberton and the organisers of the \textit{Developments of Quantitative Finance Programme}  at the Isaac Newton Institute for Mathematical Sciences in Cambridge, 2005;  Professor Goran Peskir, Doctor Pavel Gapeev and Professor Luis Alvarez and the organisers of the \textit{Symposium on  Optimal Stopping with Applications}  in Manchester, January 2006; Professor Richard Stockbridge and and the organisers of the \textit{Stochastic Filtering and Control Workshop}  at Warwick, August 2007 for helpful comments and discussions. Finally, Catherine Donnelly has provided some helpful suggestions on the text.

%\bibliography{all}

\begin{thebibliography}{38}
\providecommand{\natexlab}[1]{#1}
\providecommand{\url}[1]{\texttt{#1}}
\expandafter\ifx\csname urlstyle\endcsname\relax
  \providecommand{\doi}[1]{doi: #1}\else
  \providecommand{\doi}{doi: \begingroup \urlstyle{rm}\Url}\fi

\bibitem[Alvarez(2001)]{A_RF}
L.~H.~R. Alvarez.
\newblock Reward functionals, salvage values, and optimal stopping.
\newblock \emph{Mathematical Methods of Operations Research}, 54:\penalty0
  315--337, 2001.

\bibitem[Beibel and Lerche(1997)]{bl97}
M.~Beibel and H.~R. Lerche.
\newblock A new look at optimal stopping problems related to mathematical
  finance.
\newblock \emph{Statistica Sinica}, 7:\penalty0 93--108, 1997.

\bibitem[Beibel and Lerche(2000)]{bl00}
M.~Beibel and H.~R. Lerche.
\newblock A note on optimal stopping of regular diffusions under random
  discounting.
\newblock \emph{Theory of Probability and its Applications}, 45:\penalty0
  657--669, 2000.

\bibitem[Bensoussan and Lions(1982)]{BL_AVI}
A.~Bensoussan and J.~L. Lions.
\newblock \emph{Applications of variational inequalities in stochastic
  control.}, volume~12 of \emph{Studies in Mathematics and its Applications}.
\newblock North Holland, 1982.

\bibitem[Borodin and Salminen(2002)]{hndbkbm}
A.~N. Borodin and P.~Salminen.
\newblock \emph{Handbook of Brownian Motion - Facts and Formulae}.
\newblock Birkhauser-Verlag, 2002.

\bibitem[Breiman(1968)]{breiman}
L.~Breiman.
\newblock \emph{Probability}.
\newblock Addison-Wesley, 1968.

\bibitem[Bronstein et~al.(2005)Bronstein, Hughston, Pistorius, and
  Zervos]{BHPZ}
A.~L. Bronstein, L.~P. Hughston, M.~R. Pistorius, and M.~Zervos.
\newblock Discretionary stopping of one-dimensional {It\^{o}} diffusions with a
  staircase payoff function.
\newblock \emph{Journal of Applied Probability}, 43:\penalty0 984--996, 2005.

\bibitem[Christensen and Irle(2011)]{CI_HFTOS}
S.~Christensen and A.~Irle.
\newblock A harmonic function technique for the optimal stopping of diffusions.
\newblock \emph{Stochastics An International Journal of Probability and
  Stochastic Processes}, 83\penalty0 (4-6):\penalty0 347--363, 2011.

\bibitem[Dai et~al.(2010)Dai, Zhang, and Zhu]{DZZ_TFTRSM}
M.~Dai, Q.~Zhang, and Q.~J. Zhu.
\newblock Trend following trading under a regime switching model.
\newblock \emph{SIAM Journal on Financial Mathematics}, 1:\penalty0 780--810,
  2010.

\bibitem[Davis and Karatzas(1994)]{DavKar94}
M.~H.~A. Davis and I.~Karatzas.
\newblock A deterministic approach to optimal stopping.
\newblock In F.~P. Kelly, editor, \emph{Probability, Statistics and
  Optimisation: A Tribute to Peter Whittle}, Wiley Series in Probability and
  Mathematical Statistics, pages 455--466. Wiley, 1994.

\bibitem[Dayanik(2008)]{D_OSLDRD}
S.~Dayanik.
\newblock Optimal stopping of linear diffusions with random discounting.
\newblock \emph{Mathematics of Operations Research}, 33\penalty0 (3):\penalty0
  645--661, 2008.

\bibitem[Dayanik and Karatzas(2003)]{daykar}
S.~Dayanik and I.~Karatzas.
\newblock On the optimal stopping problem for one-dimensional diffusions.
\newblock \emph{Stochastic Processes and their Applications}, 107\penalty0
  (2):\penalty0 173--212, 2003.

\bibitem[Dixit and Pindyck(1994)]{DP94}
A.~K. Dixit and R.~S. Pindyck.
\newblock \emph{Investment under Uncertainty}.
\newblock Princeton University Press, 1994.

\bibitem[Dynkin(1963)]{D_OC}
E.~B. Dynkin.
\newblock Optimal choice of the stopping moment of a {M}arkov process.
\newblock \emph{Doklady Akademii Nauk SSSR}, 150:\penalty0 238--240, 1963.

\bibitem[El-Karoui(1979)]{sf09elk}
N.~El-Karoui.
\newblock Les aspects probabilistes du contr\^{o}le stochastique.
\newblock In P.~L. Hennequin., editor, \emph{Ecole d'Et\'e de Probabilit\'es de
  Saint-Flour IX-1979}, Lecture notes in mathematics, 876, pages 73--238.
  Springer-Verlag, 1979.

\bibitem[Feller(1952)]{feller52}
W.~Feller.
\newblock The parabolic differential equations and the associated semi-groups
  of transformations.
\newblock \emph{Annals of Mathematics}, 55\penalty0 (3):\penalty0 468--519,
  1952.

\bibitem[Guo and Shepp(2001)]{GuoSh01}
X.~Guo and L.~A. Shepp.
\newblock Some optimal stopping problems with non-trivial boundaries for
  pricing exotic options.
\newblock \emph{Journal of Applied Probability}, 38:\penalty0 647--658, 2001.

\bibitem[Guo and Tomecek(2008)]{GT_CSCOS}
X.~Guo and P.~Tomecek.
\newblock Connections between singular control and optimal switching.
\newblock \emph{SIAM Journal on Control and Optimization}, 47\penalty0
  (1):\penalty0 421--443, 2008.

\bibitem[Henderson and Hobson(2002)]{HenHob02}
V.~Henderson and D.~Hobson.
\newblock Real options with constant relative risk aversion.
\newblock \emph{Journal of Economic Dynamics and Control}, 27\penalty0
  (2):\penalty0 329--355, 2002.

\bibitem[It{\^o} and McKean(1974)]{itomck}
K.~It{\^o} and H.~P. McKean.
\newblock \emph{Diffusion Processes and their Sample Paths}.
\newblock Springer-Verlag, 1974.

\bibitem[Johnson and Zervos(2007)]{JZ1}
T.~C. Johnson and M.~Zervos.
\newblock The solution to a second order linear ordinary differential equation
  with a non-homogeneous term that is a measure.
\newblock \emph{Stochastics}, 2007.

\bibitem[Johnson and Zervos(2010)]{JZ_ESSSP}
T.~C. Johnson and M.~Zervos.
\newblock The explicit solution to a sequential switching problem with
  non-smooth data.
\newblock \emph{Stochastics}, 82\penalty0 (1):\penalty0 69--109, 2010.

\bibitem[Karatzas and Shreve(1991)]{ks1}
I.~Karatzas and S.~Shreve.
\newblock \emph{Brownian Motion and Stochastic Calculus}.
\newblock Springer-Verlag, 1991.
\newblock ISBN 0387976558.

\bibitem[Karlin(1962)]{K_OMSA}
S.~Karlin.
\newblock Stochastic models and optimal policy for selling an asset.
\newblock In K.~J. Arrow, S.~Karlin, and H.~Scarf, editors, \emph{Studies in
  Applied Probability and Management Science}. Stanford University Press, 1962.

\bibitem[Karlin and Taylor(1981)]{karlin}
S.~Karlin and H.~M. Taylor.
\newblock \emph{A Second Course in Stochastic Processes}.
\newblock Academic Press, 1981.

\bibitem[Krylov(1980)]{kryl80}
N.~V. Krylov.
\newblock \emph{Controlled diffusion processes}.
\newblock Springer-Verlag, 1980.
\newblock URL

\bibitem[Lamberton(2009)]{L_OSIRF}
D.~Lamberton.
\newblock Optimal stopping with irregular reward functions.
\newblock \emph{Stochastic Processes and their Applications}, 119\penalty0
  (10):\penalty0 3253--3284, 2009.

\bibitem[Lamberton and Zervos(2012)]{LZ_OSODD}
D.~Lamberton and M.~Zervos.
\newblock On the optimal stopping of a one-dimensional diffusion.
\newblock ArXiv e-print 1207.5491, July 2012.

\bibitem[Lempa(2010)]{L_OSTS}
J.~Lempa.
\newblock A note on optimal stopping of diffusions with a two--sided optimal
  rule.
\newblock \emph{Operations Research Letters}, 38\penalty0 (1):\penalty0 11--16,
  2010.

\bibitem[Lerche and Urusov(2007)]{LU_OSMT}
H.~R. Lerche and M.~Urusov.
\newblock Optimal stopping via measure transformation: the beibel-lerche
  approach.
\newblock \emph{Stochastics}, 79\penalty0 (3--4):\penalty0 275--291, 2007.

\bibitem[McDonald and Siegel(1986)]{McDS}
R.~McDonald and D.~Siegel.
\newblock The value of waiting to invest.
\newblock \emph{The Quarterly Journal of Economics}, 101\penalty0 (4):\penalty0
  707--728, 1986.

\bibitem[Merton(1990)]{M_CTF}
R.~C. Merton.
\newblock \emph{Continuous-time Finance}.
\newblock Blackwell, 1990.

\bibitem[Rogers and Williams(1994)]{r&w2}
L.~C.~G. Rogers and D.~Williams.
\newblock \emph{Diffusions, Markov Processes and Martingales - Volume 2: It\^o
  Calculus}.
\newblock Wiley, 2nd edition, 1994.
\newblock ISBN 0-471-91482-7.

\bibitem[R\"{u}schendorf and Urusov(2008)]{RU_COSP}
L.~R\"{u}schendorf and M.~Urusov.
\newblock On a class of optimal stopping problems for diffusions with
  discontinuous coefficients.
\newblock \emph{Annals of Applied Probability}, 18\penalty0 (3):\penalty0
  847--878, 2008.

\bibitem[Salminen(1985)]{Sal85}
P.~Salminen.
\newblock Optimal stopping of one-dimensional diffusions.
\newblock \emph{Mathematische Nachrichten}, 124:\penalty0 85--101, 1985.

\bibitem[Shiryaev(1978)]{shir78}
A.~N. Shiryaev.
\newblock \emph{Optimal stopping rules}.
\newblock Springer-Verlag, 1978.

\bibitem[Shreve(2004)]{S_SCF2}
S.~E. Shreve.
\newblock \emph{Stochastic Calculus for Finance II: Continuous-Time Models}.
\newblock Springer, 2004.

\bibitem[Trigeorgis(1996)]{T96}
L.~Trigeorgis.
\newblock \emph{Real Options: Managerial Flexibility and Strategy in Resource
  Allocation}.
\newblock MIT Press, 1996.

\end{thebibliography}

\end{document}